\documentclass{emulateapj}
\usepackage{natbib}
\usepackage[caption=false]{subfig}

\begin{document}
\title{Does the Debris Disk around HD 32297 Contain Cometary Grains?\footnotemark[*]\footnotemark[+]}
\footnotetext[*]{Based on observations made at the Large Binocular Telescope (LBT). The LBT is an international collaboration among institutions in the United States, Italy and Germany. LBT Corporation partners are: The University of Arizona on behalf of the Arizona university system; Istituto Nazionale di Astrosica, Italy; LBT Beteiligungsgesellschaft, Germany, representing the Max-Planck Society, the Astrophysical Institute Potsdam, and Heidelberg University; The Ohio State University, and The Research Corporation, on behalf of The University of Notre Dame, University of Minnesota and University of Virginia.}
\footnotetext[+]{Based on observations made using the Large Binocular Telescope Interferometer (LBTI). LBTI is funded by the National Aeronautics and Space Administration as part of its Exoplanet Exploration program.}
\author{Timothy J. Rodigas\altaffilmark{1,2}, John H. Debes\altaffilmark{3}, Philip M. Hinz\altaffilmark{1}, Eric E. Mamajek\altaffilmark{4}, Mark J. Pecaut\altaffilmark{4,5}, Thayne Currie\altaffilmark{6}, Vanessa Bailey\altaffilmark{1}, Denis Defrere\altaffilmark{1}, Robert J. De Rosa\altaffilmark{7}, John M. Hill\altaffilmark{8}, Jarron Leisenring\altaffilmark{1}, Glenn Schneider\altaffilmark{1}, Andrew J. Skemer\altaffilmark{1}, Michael Skrutskie\altaffilmark{9}, Vidhya Vaitheeswaran\altaffilmark{1}, Kimberly Ward-Duong\altaffilmark{7}}

\altaffiltext{1}{Steward Observatory, The University of Arizona, 933 N. Cherry Ave., Tucson, AZ 85721, USA; email: rodigas@as.arizona.edu}
\altaffiltext{2}{Carnegie Postdoctoral Fellow; Department of Terrestrial Magnetism, Carnegie Institute of Washington, 5241 Broad Branch Road, NW, Washington, DC 20015, USA}
\altaffiltext{3}{Space Telescope Science Institute, Baltimore, MD 21218, USA}
\altaffiltext{4}{Department of Physics and Astronomy, University of Rochester, Rochester, NY 14627-0171, USA}
\altaffiltext{5}{Rockhurst University, 1100 Rockhurst Rd, Kansas City, MO 64110, USA}
\altaffiltext{6}{University of Toronto, 50 St George St., Toronto, ON M5S 1A1, Canada}
\altaffiltext{7}{School of Earth and Space Exploration, Arizona State University, PO Box 871404, Tempe, AZ 85287-1404, USA}
\altaffiltext{8}{Large Binocular Telescope Observatory, University of Arizona, Tucson, AZ 85721, USA}
\altaffiltext{9}{University of Virginia, Department of Astronomy, 530 McCormick Road, Charlottesville, VA  22903, USA}

\newcommand{\about}{$\sim$~}
\newcommand{\mj}{M$_{J}$}
\newcommand{\degrees}{$^{\circ}$}
\newcommand{\arcseconds}{$^{\prime \prime}$}
\newcommand{\asec}{$\arcsec$}
\newcommand{\fasec}{$\farcs$}
\newcommand{\lprime}{$L^{\prime}$}
\newcommand{\ks}{$Ks$~}
\newcommand{\mjyasec}{mJy/arcsecond$^{2}$}
\newcommand{\microns}{$\mu$m}

\shortauthors{Rodigas et al.}

\begin{abstract}
We present an adaptive optics imaging detection of the HD 32297 debris disk at \lprime ~(3.8 \microns) obtained with the LBTI/LMIRcam infrared instrument at the LBT. The disk is detected at signal-to-noise per resolution element (SNRE) \about 3-7.5 from \about 0\fasec 3-1\fasec 1 (30-120 AU). The disk at \lprime ~is bowed, as was seen at shorter wavelengths. This likely indicates the disk is not perfectly edge-on and contains highly forward scattering grains. Interior to \about 50 AU, the surface brightness at \lprime ~rises sharply on both sides of the disk, which was also previously seen at \ks band. This evidence together points to the disk containing a second inner component located at $\lesssim$ 50 AU. Comparing the color of the outer (50 $< r$/AU $< 120$) portion of the disk at \lprime ~with archival HST/NICMOS images of the disk at 1-2 \microns ~allows us to test the recently proposed cometary grains model of \cite{donaldson32297}. We find that the model fails to match the disk's surface brightness and spectrum simultaneously (reduced chi-square = 17.9). When we modify the density distribution of the model disk, we obtain a better overall fit (reduced chi-square = 2.9). The best fit to all of the data is a pure water ice model (reduced chi-square = 1.06), but additional resolved imaging at 3.1 \microns ~is necessary to constrain how much (if any) water ice exists in the disk, which can then help refine the originally proposed cometary grains model. 
\end{abstract}

\section{Introduction}
Debris disks, which are thought to be continually replenished by collisions between large planetesimals \citep{wyatt08}, can point to interesting planets in several ways: with warps and/or gaps \citep{betapic}, sharp edges \citep{hr4796schneider,chiang}, and with their specific dust grain compositions. Since many outer solar system bodies contain copious amounts of water ice and organic materials, finding other debris disk systems that contain water ice and/or organic materials \citep{4796organics} would point to planetary systems that might contain the ingredients necessary for Earth-like life. Therefore constraining the dust grain compositions in debris disks is crucial.


Narrow and broadband scattered light imaging is a particularly powerful tool for constraining composition because it can substitute for spectra that would otherwise be too difficult to obtain. The wavelength range between 1-5 \microns, in particular, contains strong absorption features for water ice and organics like tholins (both near 3.1 \microns; \cite{inoue,phoebe}). Imaging at these wavelengths from the ground using adaptive optics (AO) also offers high Strehl ratios, allowing for more precise characterization of faint extended sources close to their host stars.

Obtaining high signal-to-noise (S/N) detections of faint debris disks at these wavelengths from the ground is challenging due to the bright thermal background of Earth's atmosphere, as well as the warm glowing surfaces in the optical path of the telescope. An AO system that suppresses unwanted thermal noise is necessary to overcome these obstacles. The Large Binocular Telescope (LBT), combined with the Large Binocular Telescope Interferometer (LBTI, \cite{lbti}), is one such system. The LBT AO system \citep{lbtao} consists of two secondary mirrors (one for each primary) that can each operate with up to 400 modes of correction, resulting in very high Strehl ratios (\about 70-80$\%$ at $H$ band, \about 90$\%$ at \ks band, and $>$ 90$\%$ at longer wavelengths). This equates to very high-contrast, high-sensitivity imaging capabilities, allowing detections of planets and debris disks that were previously too difficult.

HD 32297 is a young A star located 112 pc away \citep{updatedhip} surrounded by a bright edge-on debris disk. The disk has recently been resolved at \ks ~band (2.15 \microns) by \cite{currie32297}, \cite{boc32297}, and \cite{esposito32297}. The disk has also been detected in the far-infrared (FIR) by Herschel (\citealt{donaldson32297}, hereafter D13), who modeled the disk as consisting of porous cometary grains (silicates, carbonaceous material, water ice). 

To further test this model, we have obtained a high S/N image of the disk at \lprime ~(3.8 \microns) with LBTI/LMIRcam. Using this new image, along with archival HST/NICMOS images of the disk at 1-2 \microns ~from \cite{debes32297}\footnote{While resolved images of the disk have also been obtained from the ground at these wavelengths, the HST data are preferred because they do not suffer from the biases inherent in ADI/PCA data reduction.}, we determine how well the D13 cometary grains model matches the scattered light of the disk at 1-4 \microns.

In Section 2 we describe the observations and data reduction. In Section 3 we present our results on the disk's surface brightness (SB) from 1-4 \microns, analysis of the disk's morphology, ~and our detection limits on planets in the system. In Section 4 we present our modeling of the disk. In Section 5 we discuss the implications of our results on the disk's structure and composition, and in Section 6 we summarize and conclude. 

\section{Observations and Data Reduction}
\subsection{Observations}
We observed HD 32297 on the night of UT November 4 2012 at the LBT on Mt. Graham in Arizona. We used LBTI/LMIRcam and observed at \lprime ~(3.8 \microns). LMIRcam has a field of view (FOV) of \about 11\asec ~on a side and a plate scale of 0\fasec 0107/pixel. Skies were clear during the observations, and the seeing was \about 1\fasec 3 throughout. Observations were made in angular differential imaging mode (ADI; \cite{adi}), and no coronagraphs were used. We used the single left primary mirror with its deformable secondary mirror operating at 400 modes for the AO correction. To increase nodding efficiency, we chopped an internal mirror (rather than nodding the entire telescope) by several arcseconds every few minutes to obtain images of the sky background. While this method did dramatically increase efficiency (to \about 85$\%$), neglecting to move the telescope resulted in a residual ``patchy" background that remained even after sky subtraction. This was due to the difference in optical path through the instrument, such that the chop positions were not perfectly matched. Ultimately we removed this unwanted background by unsharp-masking all images, which results in self-subtraction of the debris disk; we account for and mitigate this effect via insertion of artificial disks, which we discuss in Section \ref{sec:corrections}.

We obtained 759 images of HD 32297 in correlated double sampling (CDS) mode, so that each image cube consisted of 15 coadded minimum exposure (0.029 s) images containing the unsaturated star (for photometric comparison), and 15 coadded 0.99 s science exposure images with the core of the star saturated out to 0\fasec 1. After filtering out images taken while the AO loop was open\footnote{Filtering out images of poor quality can be accomplished in several ways: the fits headers contain keywords relating to the status of the AO loop, which a user can check in the data reduction; the raw images themselves can be examined by the user to check the appearance of the PSF (pointy vs. blurry); or the user can check the log of the observations, which should denote when/where problems with the AO occurred. We used the first and third options to filter out open loop data on HD 32297.}, the final dataset consisted of 726 images, resulting in 2.99 hours of continuous integration (not including the minimum exposure photometric images). Throughout the observations, which began just after the star's transit, the FOV rotated by 50.84\degrees, enabling the star itself to act as the point spread function (PSF) reference for subtraction during data reduction. 

\subsection{Data reduction}
All data reduction discussed below was performed with custom Matlab scripts. We first divided each science exposure image by the number of coadds (15) and integration time (0.99 s) to obtain units of counts/s for each pixel. Next we corrected for bad pixels and subtracted opposite chop beam images of the star to remove detector artifacts and the sky background, resulting in flat images with \about 0 background counts/s, except for the ``patchy" regions of higher sky noise. We determined the sub-pixel location of the star in each sky-subtracted image by calculating the center of light inside a 0\fasec 5 aperture centered on the approximate location of the star.\footnote{Since the PSF is only saturated out to 0\fasec 1, it contributes very little to the center of light calculation; we have demonstrated sub-pixel accuracy using this method in \cite{mehd15115}.} We then registered each image so that the star's location was at the exact center of each image. We binned each image by a factor of 2 to ease the computational load required in processing 726 images, which is not a problem because the PSF is still oversampled by more than a factor of 4. To reduce the level of the patchy sky background, for each image we subtracted a 15 pixel (0\fasec 32) by 15 pixel box median-smoothed image from itself. The unsaturated minimum exposure images were reduced as detailed above, except for this last step of unsharp masking, since the majority of the star's flux does not overlap with the background patches. It is also not necessary to unsharp mask the unsaturated photometric images because, as will be discussed in Section \ref{sec:corrections}, we compute the fully-corrected disk flux that accounts for the unsharp masking and other biases inherent in the data reduction. 

During the observing run, the LMIRcam detector suffered from ``S-shaped" non-linearity (which has since been corrected). This had the effect of artificially inflating raw counts per pixel at long exposures relative to short exposures, complicating photometric comparisons. Using a linearity curve constructed from images taken throughout the observing run, we multiplied each reduced image's pixels by the corresponding linearity correction factors. We verified the effectiveness of the linearity correction by comparing the total flux within an annulus centered on the star between 0\fasec 1-0\fasec 2 for the unsaturated (linear) final image and the final (linearity-corrected) long exposure image; the median counts/s for the two images agreed to within \about 4$\%$, verifying the linearity correction.

As a first check on the efficacy of the steps described above, we performed classical ADI subtraction by subtracting a median-combined master PSF image from all the images, and then derotating the images by their parallactic angle at the time of the exposure. The resulting image revealed edge-on disk structure at the expected position angle (PA) of \about 47\degrees \citep{debes32297,mawet32297}.

\begin{figure*}[t]
\centering
\subfloat[]{\label{fig:finaldisk}\includegraphics[scale=0.43]{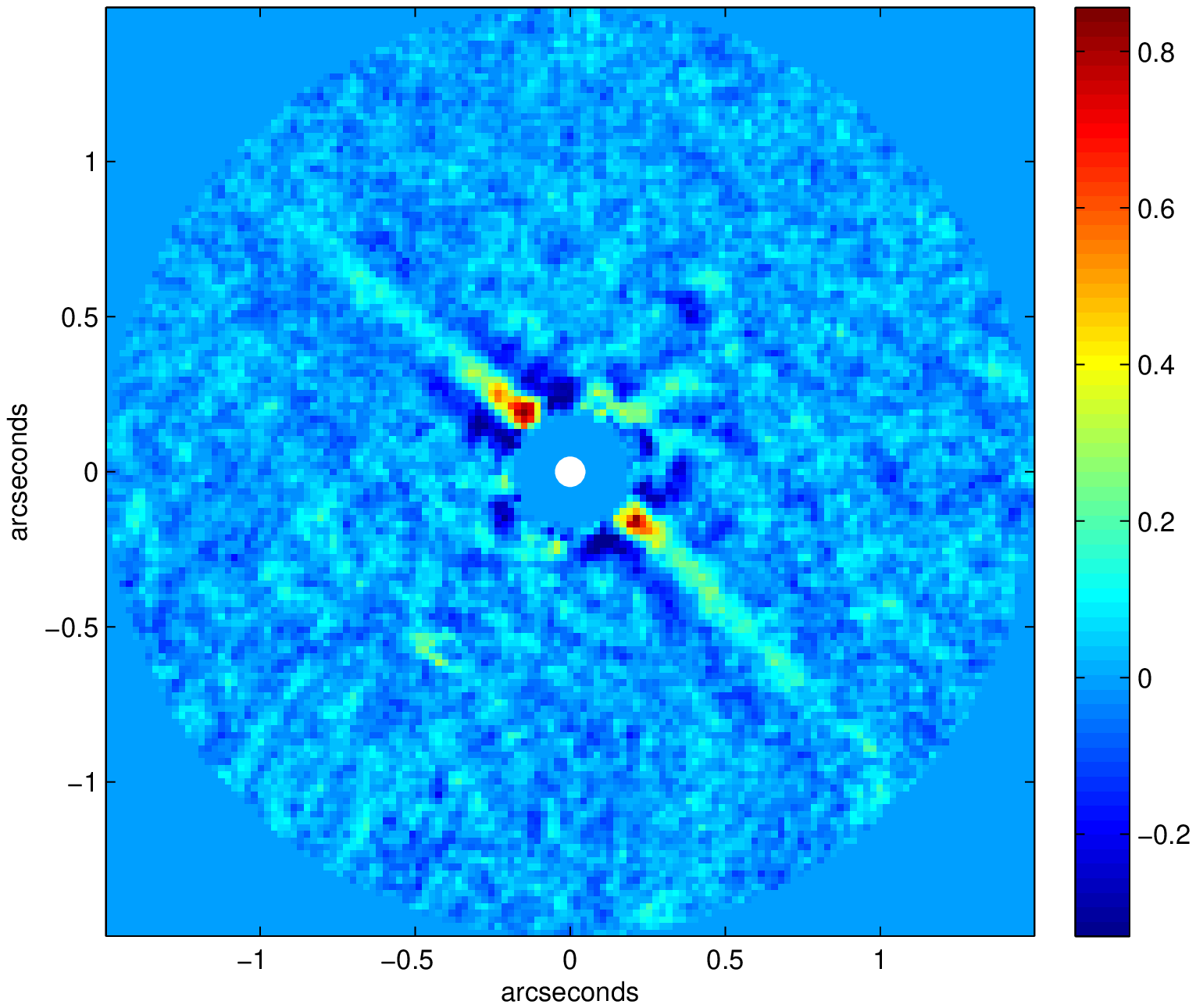}} 
\subfloat[]{\label{fig:finaldisksnre}\includegraphics[scale=0.43]{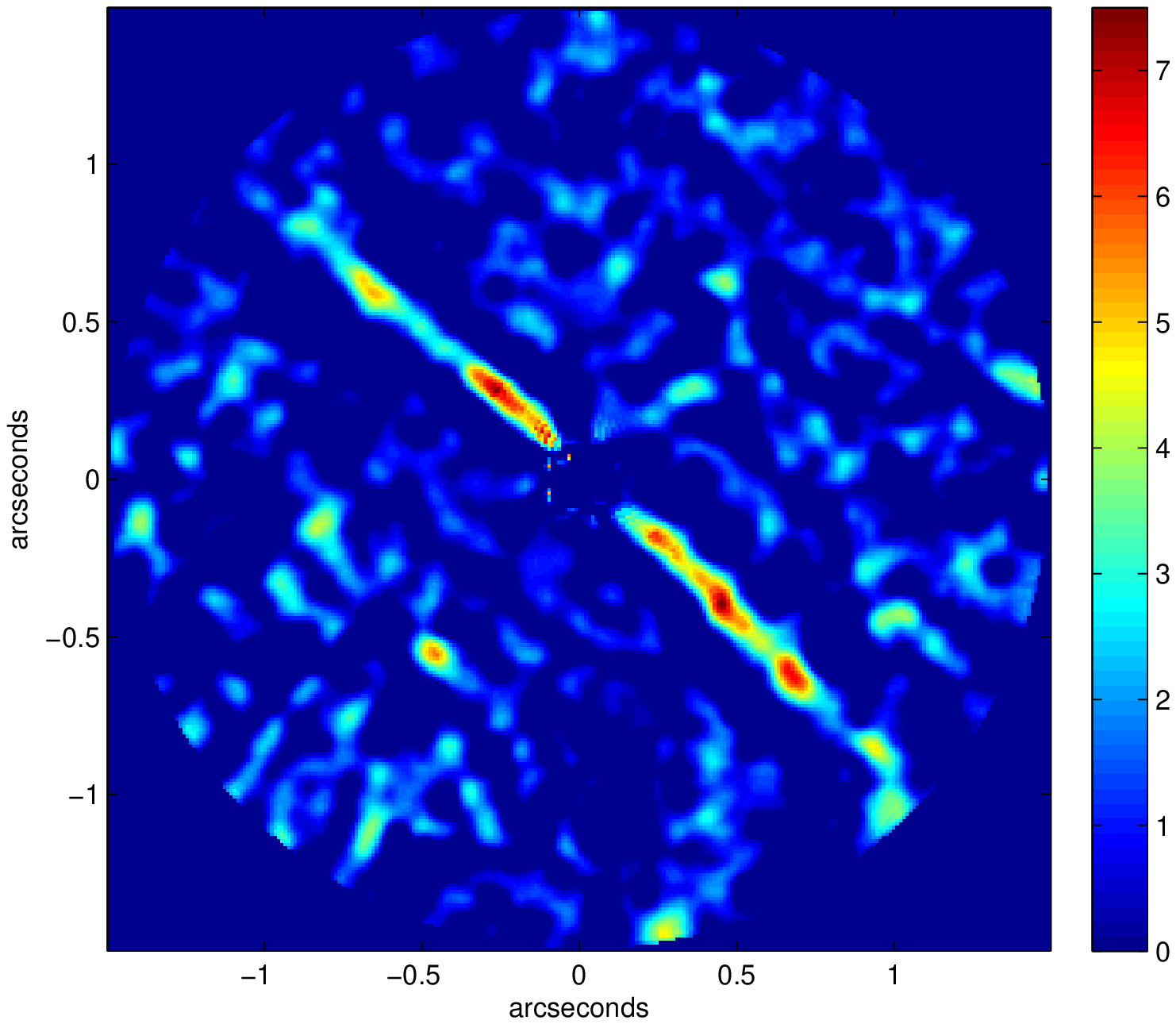}}
\caption{\textit{Left}: Final reduced \lprime ~image of the HD 32297 debris disk, in units of detector counts/s, with North-up, East-left. The white dot marks the location of the star and represents the size of a resolution element at \lprime. A 0\fasec 2 radius mask has been added in post-processing. The southwest side of the disk is \about 0.5 magnitudes/arcsecond$^{2}$ brighter than the northeast side from 0\fasec 5-0\fasec 8 (56-90 AU), which was also seen at \ks band \citep{currie32297}. However there is no brightness asymmetry at the location of the mm peak first identified by \cite{maness32297} and later seen at \ks band by \cite{currie32297}. \textit{Right}: SNRE map of the final image. Both sides of the disk are detected from \about 0\fasec 3-1\fasec 1 (30-120 AU) at SNRE \about 3-7.5.}
\end{figure*}

To obtain the highest possible S/N detection of HD 32297's debris disk, we reduced the images using principal component analysis (PCA, \cite{pca}). PCA has recently been shown to produce equal-to or higher S/N detections of planets and disks \citep{hip79977,bonnefoybetapic,bocbetapic,pca,tiffany2} than LOCI \citep{loci}, though this may be a result of LOCI's tunable parameters not being optimized correctly (C. Marois, private communication). We did also reduce the data using conventional LOCI algorithms \citep{mehd15115,currie32297,thalmannhr4796} but found that the gain in computational speed for PCA with no loss in S/N warranted the ultimate preference of PCA over LOCI for this dataset. 

We followed the prescription outlined in \cite{pca}, with the main tunable parameter being $K$, the number of modes to use for a given reduction. Increasing $K$ reduces the noise in the final image but also suppresses the flux from the disk; therefore this parameter must be optimized. After examining the average S/N per resolution element (SNRE)\footnote{As in \cite{mehd15115}, the SNRE map is calculated by convolving the final image by a Gaussian with full-width half-maximum = 0\fasec 094, constructing a noise image by computing and storing the standard deviation in concentric annuli of width = 1 pixel around the star, then dividing the final Gaussian-smoothed image by this noise image.} over the disk's spatial extent for varying values of $K$, we determined the optimal number of modes to be $K = 3$ (out of a possible 726). We then fed all the images through our PCA pipeline, rotated the images by their corresponding PAs clockwise to obtain North-up East-left, and combined the images using a mean with sigma clipping. Fig. \ref{fig:finaldisk} shows this final image, and Fig. \ref{fig:finaldisksnre} shows the corresponding SNRE map. We detect the disk from \about 0\fasec 3-1\fasec 1 (30-120 AU) at SNRE \about 3-7.5, which is a significant improvement from our previous \lprime ~imaging of HD 15115 \citep{mehd15115}. The detection of the disk at high S/N allows us to more precisely measure its SB, which in turn results in better constraints on the composition and size of the dust grains producing the observed scattered light.

\section{Results}
\subsection{Calculation of Uncertainties}
For all the images analyzed in this study, the uncertainties are dominated by residuals left over from PSF subtraction. Specifically, both the HST and LBT data suffer from azimuthal and radial residual structures. For the LBT data, we computed the standard deviation of the equivalent SB measurements all around the star (excluding the disk). For the HST data, we calculated the errors in two ways: by computing the equivalent SB measurements 90\degrees away from the real disk; and by computing the standard deviation of the equivalent SB measurments all around the star (excluding the disk). We found that both methods resulted in comparable errors, therefore we chose to use the second method since this method was also used for computing the errors on the LBT data. 

\subsection{Surface Brightness Profiles}
\label{sec:sbmeasures}
\begin{figure*}[t]
\centering
\subfloat[]{\label{fig:sbeast}\includegraphics[scale=0.4]{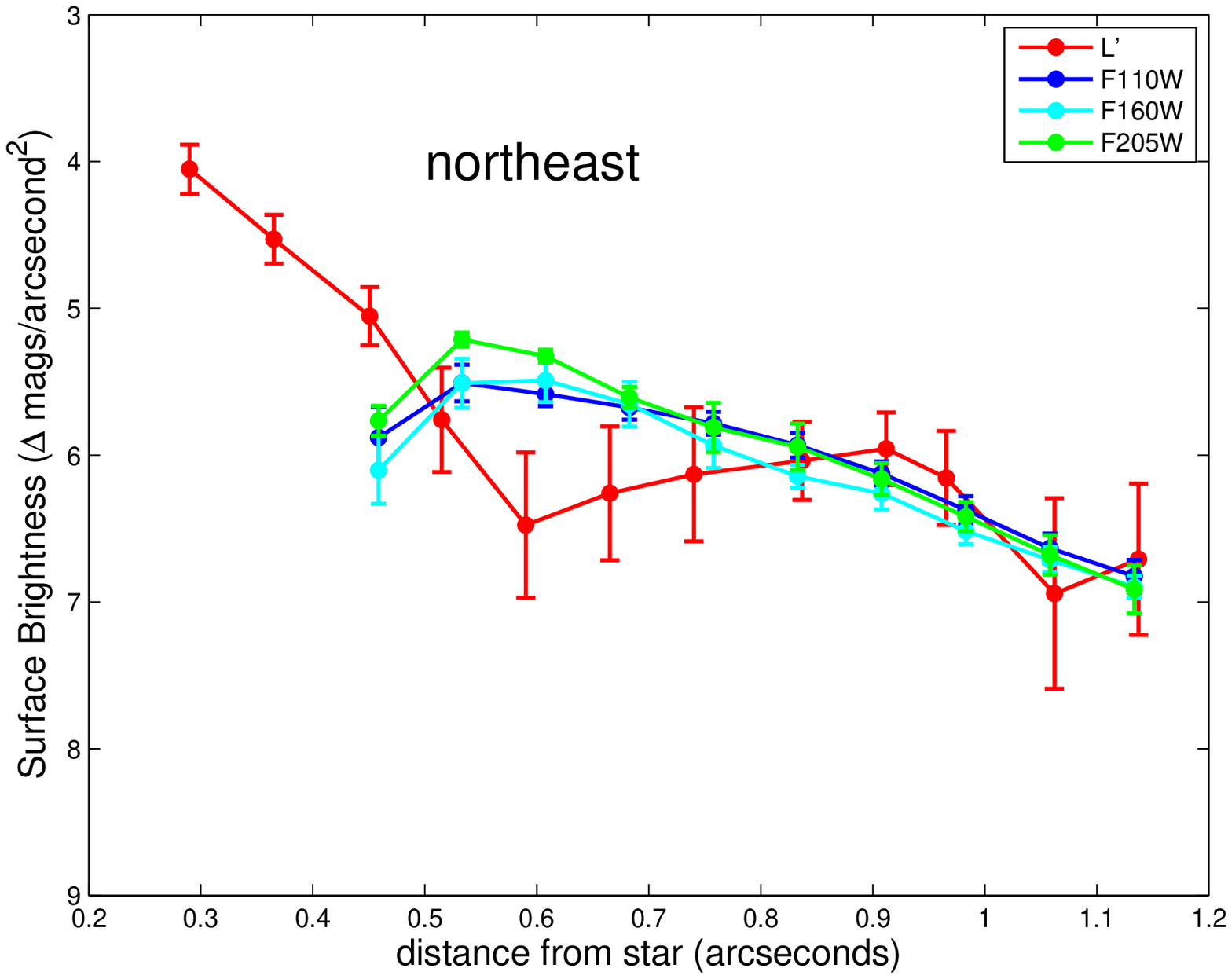}} 
\subfloat[]{\label{fig:sbwest}\includegraphics[scale=0.4]{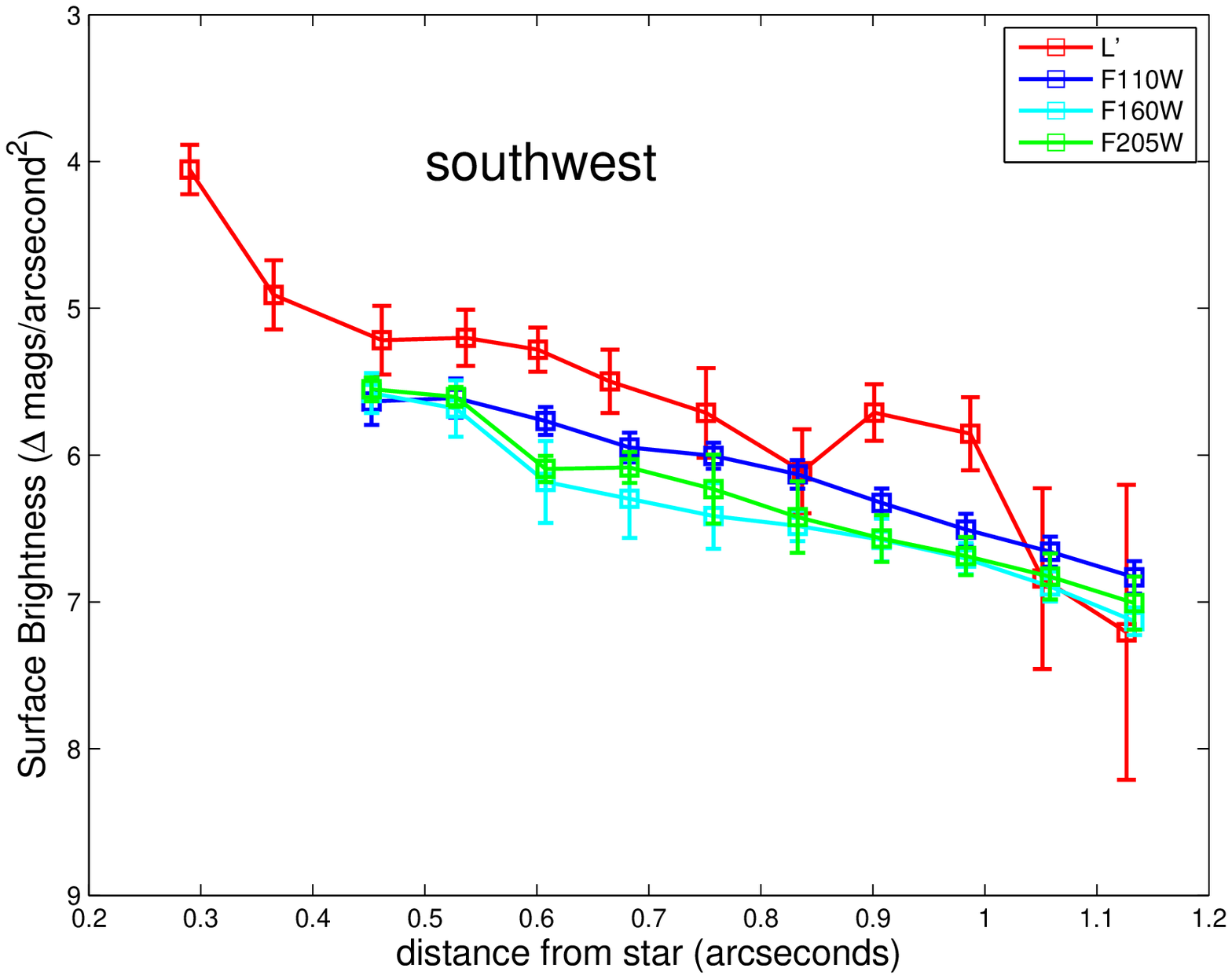}} 
\caption{\textit{Left}: SB profiles for the HD 32297 debris disk for the 1-2 \microns ~HST/NICMOS and 3.8 \microns ~LBTI data for the northeastern lobe. \textit{Right}: The same except for the southwestern lobe of the disk. Interestingly, the SB asymmetry at 0\fasec 5-0\fasec 8 (56-90 AU) first noticed by \cite{currie32297} is also evident in the \lprime ~data. Inward of 0\fasec 5, the disk's SB declines more steeply than at larger stellocentric separations, and we do not find evidence for an asymmetry near 0\fasec 4 (45 AU) that was identified as a mm peak by \cite{maness32297}. Exterior to 0\fasec 8 (90 AU), the HST 1-2 \microns ~data are generally consistent with the \lprime ~data within the uncertainties.}
\label{fig:sbplots}
\end{figure*}

\begin{table*}[t]
\centering
\caption{Surface Brightness Profile Power-law Indices}
\label{tab:indices}
\begin{tabular}{c|c c c c}
 & 3.8 \microns & 2.05 \microns & 1.6 \microns & 1.1 \microns  \\
\hline
northeast outer ($0\fasec 5 < r < 1\fasec 1$) & -1.5 (-2.45, -0.55)$^{*}$ & -1.83 (-2.12, -1.53) & -1.56 (-1.99, -1.13) & -1.33 (-1.73, -0.93) \\
southwest outer ($0\fasec 5 < r < 1\fasec 1$) & -1.3 (-1.86, -0.72) & -1.56 (-1.88, -1.24) & -1.55 (-1.92, -1.17) & -1.32 (-1.52, -1.11) \\
northeast inner ($r < 0\fasec 5$) & -2.45 (-3.36, -1.53) & -- & -- & --  \\
southwest inner ($r < 0\fasec 5$) & -2.72 (-10.74, 5.3) & -- & -- & --  \\
\hline 
\multicolumn{5}{l}{$^{*}$ These and all other parenthetical values denote 95$\%$ confidence bounds.} \\
\end{tabular} 
\end{table*}

In addition to the image of the disk at \lprime, we also reanalyzed archival, reduced images of the disk at \about 1.1, 1.6, and 2.05 \microns ~from HST/NICMOS presented in \cite{debes32297}.\footnote{We refer the reader to \cite{debes32297} for these images and descriptions of how they were reduced.} For each image, we rotated the disk by 47\degrees counterclockwise so that its midplane was approximately horizontal in the image. We measured the SB in all the HST/NICMOS images by taking the median value in 3 pixel (0\fasec 2262) by 3 pixel boxes. For the F110W (1.1 \microns) images, these boxes were centered on the brightest pixel at each specific pixel distance from the star (0\fasec 45-1\fasec 1). For the F160W (1.6 \microns) and the F205W (2.05 \microns) images, the boxes were centered on the same pixel locations as were used in the F110W image. We converted the image counts/s to \mjyasec ~using the reported photometric conversions given on the Space Telescope Science Institute website and applied the appropriate correction factors to the data (see the Appendix for a detailed description of these correction factors).

For the final \lprime ~image, we calculated the median counts/s in a 5 pixel (0\fasec 106) by 5 pixel box centered on the brightest pixel at each horizontal distance from 0\fasec 45-1\fasec1 from the star, and divided this number by the plate scale of the binned images (0\fasec 0212) squared. We used a smaller aperture for the \lprime ~data than for the HST/NICMOS data due to the disk appearing thinner (FWHM \about 0\fasec 1 $\approx \lambda$/D) at \lprime, and to avoid the large negative residuals above and below the disk close to the star. We converted these values to \mjyasec ~using the total measured flux in the unsaturated photometric image of HD 32297 and applied the appropriate correction factors to the data (see the Appendix for a detailed description of these correction factors).

Fig. \ref{fig:sbplots} shows the final, corrected SB of the disk at 1-2 \microns ~and at 3.8 \microns. The star's magnitude at 1-2 \microns ~(7.7) and \lprime ~(7.59) has been subtracted from the disk magnitude/arcsecond$^{2}$ to yield the intrinsic disk color at each wavelength. We find that the SB profile is asymmetric from \about 0\fasec 5-0\fasec 8 (56-90 AU) at \lprime, in agreement with the asymmetry at \ks band reported by \cite{currie32297} and \cite{esposito32297}. We do not find evidence for asymmetry interior to 0\fasec 4 (45 AU), however, with no evidence for the bright spot identified as a mm peak by \cite{maness32297} and also seen at \ks band by \cite{currie32297}. Exterior to 0\fasec 8 (90 AU), both sides of the disk are \about equal SB at \lprime. The HST data are generally consistent from 1-2 \microns, within the uncertainties, except for interior to 0\fasec 8 (90 AU) where the northeastern lobe appears to be brighter than the southwestern lobe. 

We fit the SB profiles at each wavelength to functions with power-law form (see Table \ref{tab:indices}). We find that the indices are all within 2$\sigma$ of each other for the outer portion of the disk ($0\fasec 5 < r < 1\fasec 1$), with most values between \about -1.3 and -1.5. Interior to 0\fasec 4, the disk is not detected at 1-2 \microns ~with HST/NICMOS but is detected at \lprime. At these close distances, the disk SB clearly falls faster, declining like \about $r^{-2.6}$. We do not compare our reported power-law indices to indices reported in other works measured farther from the star because the disk is thought to have a break in the SB distribution near 110 AU (1 \asec; \citealt{boc32297,currie32297}). 

\subsection{Midplane Offset Measurements}
\begin{figure*}[t]
\centering
\subfloat[]{\label{fig:PAeast}\includegraphics[scale=0.4]{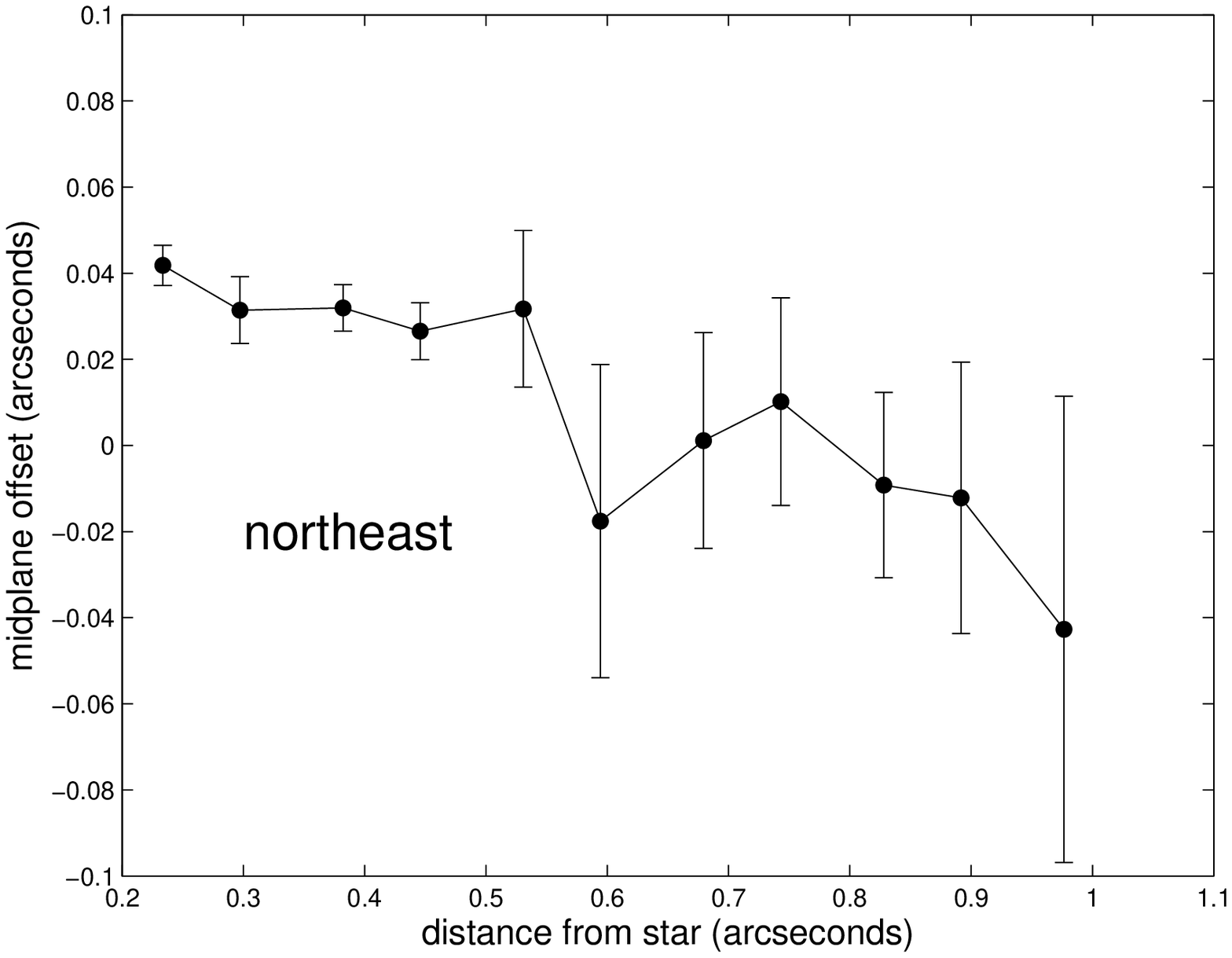}} 
\subfloat[]{\label{fig:PAwest}\includegraphics[scale=0.4]{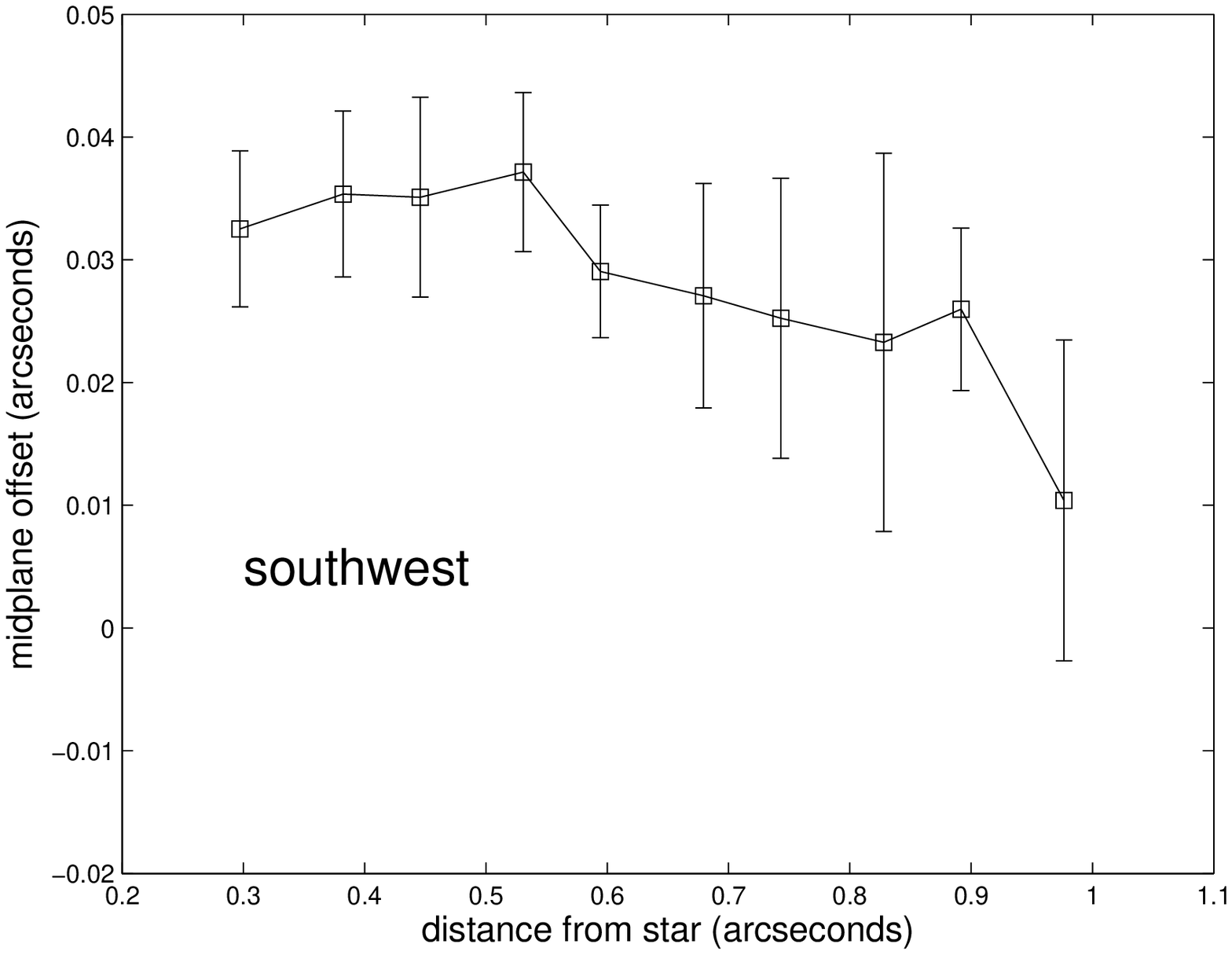}} \\
\subfloat[]{\label{fig:KsPAeast}\includegraphics[scale=0.4]{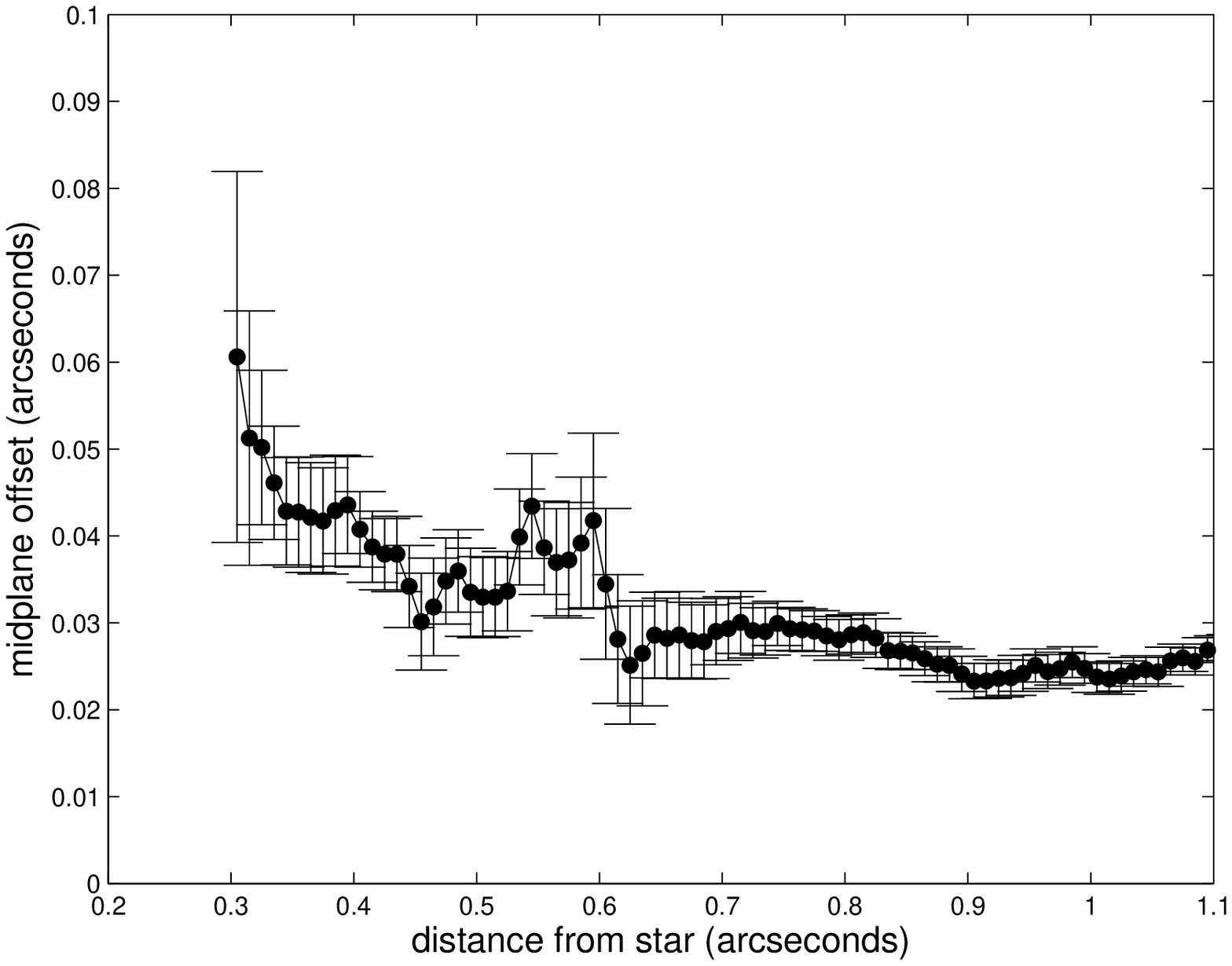}} 
\subfloat[]{\label{fig:KsPAwest}\includegraphics[scale=0.4]{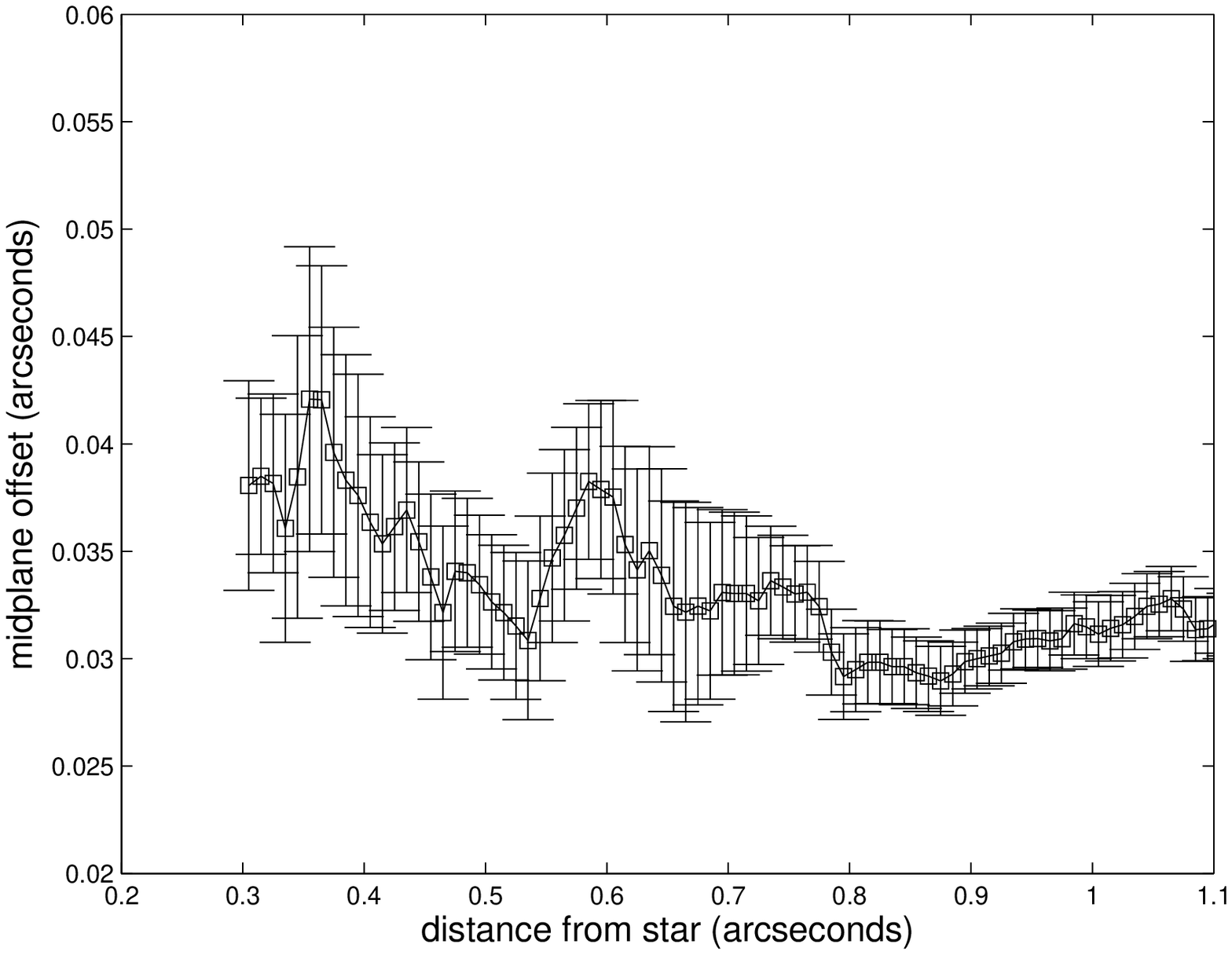}}
\caption{\textit{Top left}: Disk midplane offset as a function of distance from the star at 3.8 \microns ~for the northeastern lobe. \textit{Top right}: The same except for the southwestern lobe of the disk. \textit{Bottom row}: the same as the top row, except the offsets are measured on the \ks band data from \cite{currie32297}. The offset from the midplane increases closer to the star at both \lprime ~and \ks band, indicating a bow-shaped disk. The offsets peak at \about 0\fasec 04-0\fasec 06 (4.5-7 AU).}
\label{fig:PAs}
\end{figure*}

\cite{currie32297} measured the PA of the HD 32297 debris disk as a function of separation from the star at \ks band and found that the disk was bowed close to the star. Similar bowing was reported by \cite{boc32297} and \cite{esposito32297}. To test whether the bow shape is seen at \lprime, we measured the offset of the disk relative to the midplane as a function of distance from the star. We measure the midplane offset, as opposed to the disk's PA, because the former is a more intuitive indicator of a bow-shaped disk. The offsets were measured in manners analogous to those described in \cite{mehd15115} and \cite{currie32297}. Fig. \ref{fig:PAs} shows these offsets for the northeastern and southwestern lobes, along with the midplane offsets for the \ks band data from \cite{currie32297} for reference. The disk is clearly bowed at \lprime, with the offsets increasing closer to the star on both sides of the disk to a peak value of \about 0\fasec 04 (4.5 AU). This is comparable to the peak midplane offset reported by \cite{esposito32297} and agrees with the peak offsets in the \ks band data (Fig. \ref{fig:KsPAeast} and Fig. \ref{fig:KsPAwest}).

\subsection{Revised Spectral Classification, Luminosity, Mass, and Age of HD 32297}
\begin{figure}[h]
\centering
\subfloat[]{\label{fig:agespec}\includegraphics[scale=0.25]{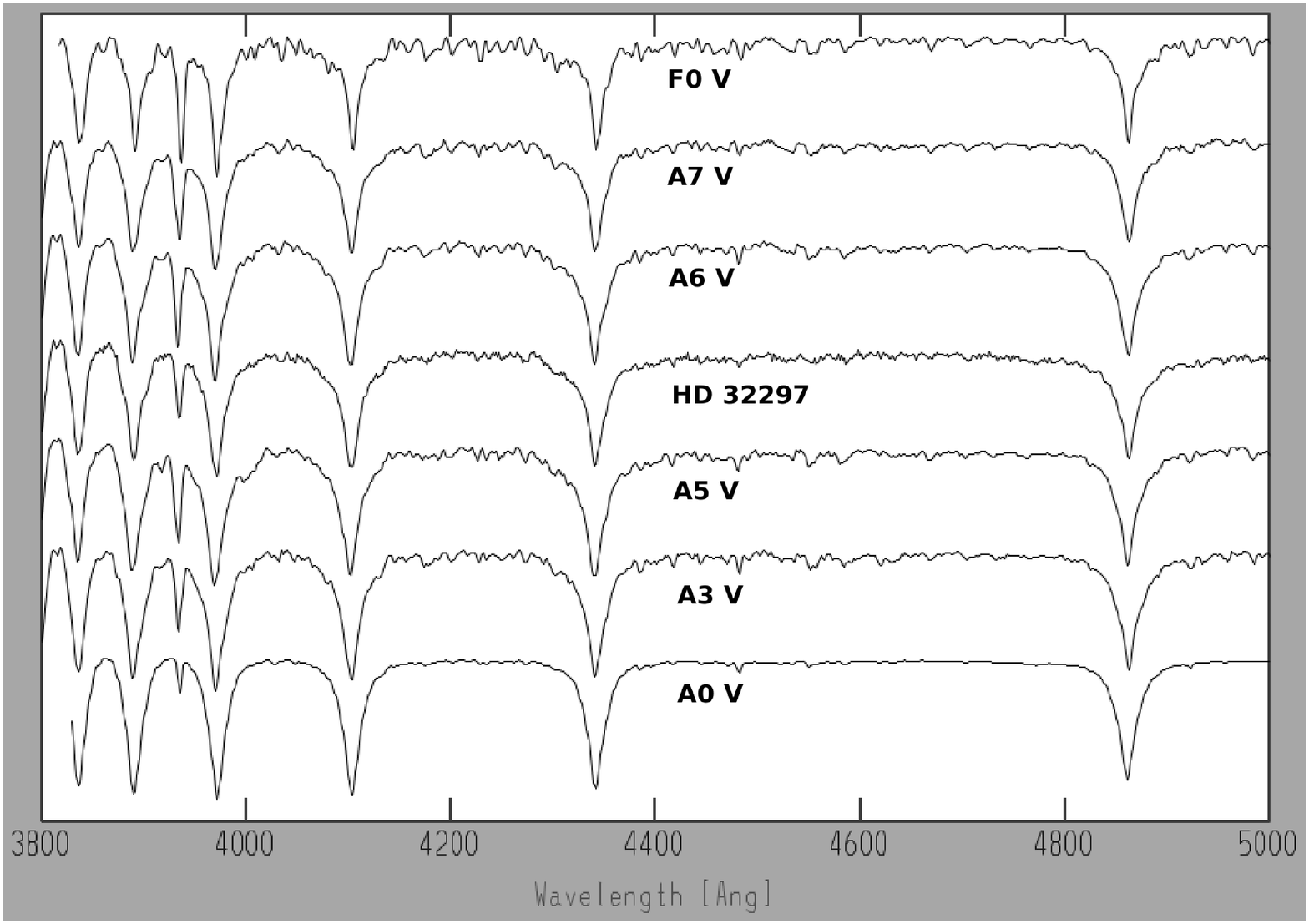}} \\
\subfloat[]{\label{fig:age}\includegraphics[scale=0.35]{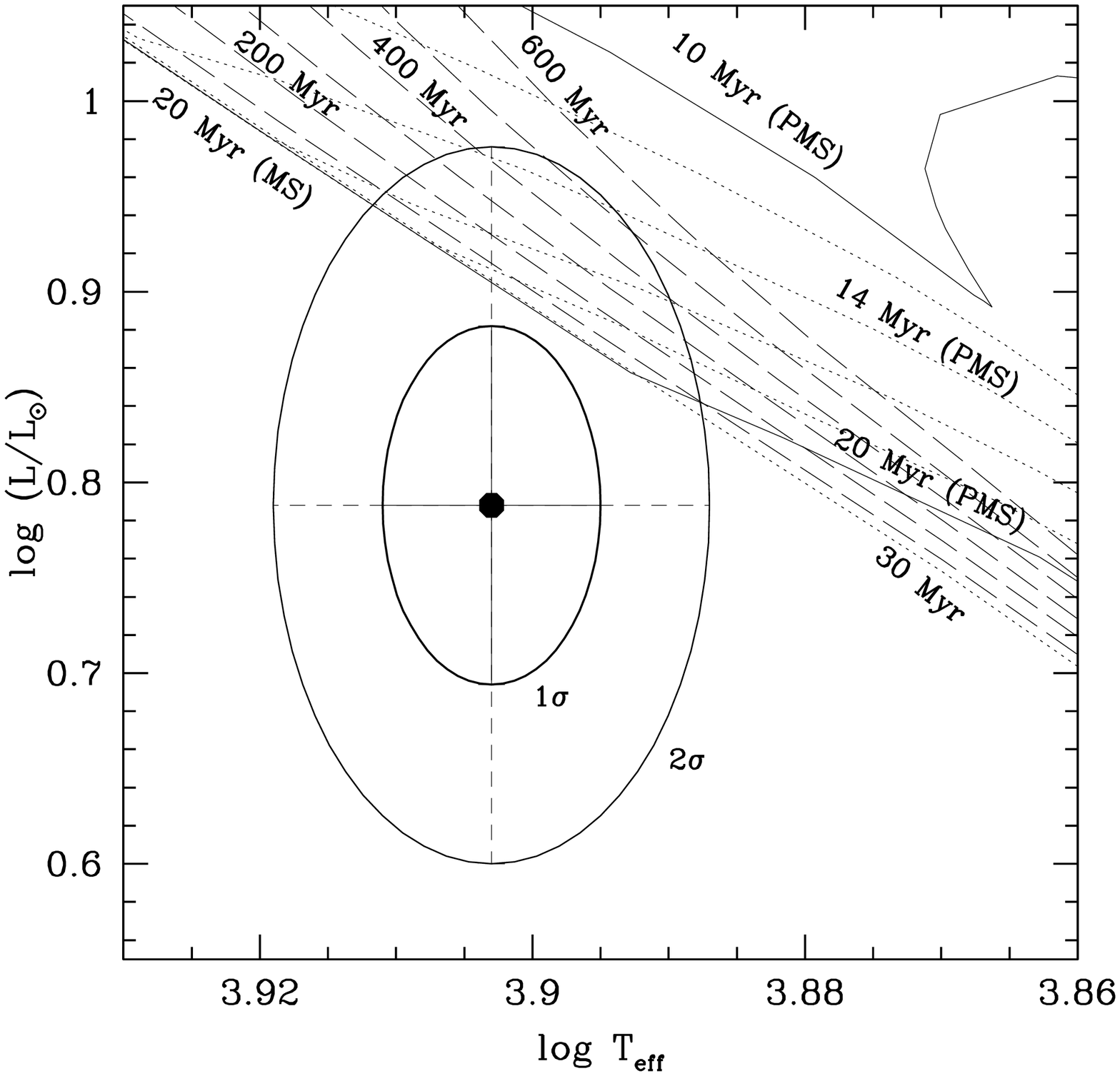}} 
\caption{\textit{Top}: Visible spectrum of HD 32297, used for determining its spectral type, which we classify as A6V. \textit{Bottom}: HD 32297's position on an HR diagram with pre-main sequence and main sequence isochrones from \citet{bressantracks} for an assumed protosolar composition with Helium mass fraction Y = 0.27 and metal mass fraction Z = 0.017. The star is likely older than \about 15 Myr and younger than \about 500 Myr. In this study, we take the age of the star to be 100 Myr.}
\end{figure}

Estimating the masses of exoplanets detected via direct imaging requires atmospheric models, which depend heavily on the age of the planet and therefore on the host star. The age of HD 32297, like many stars, is poorly constrained. Estimates of the star's spectral type, which can be used to constrain its age, have ranged from A0 \citep{32297A0} to A5 \citep{32297A5}. Based on an archival optical spectrum of the star taken on 2 February 2006 with the 300 line grating of the FAST spectrograph on the Tillinghast telescope\footnote{see FAST archive at http://tdc-www.harvard.edu/cgi-bin/arc/fsearch},
and comparison of the spectrum to a dense grid of MK standard stars using the python tool ``sptool"\footnote{http://rumtph.org/pecaut/sptool/}, we estimate the star's spectral type to be kA3hA6mA6 V (see Fig. \ref{fig:agespec}).\footnote{Each of the lower-case prefix letters corresponds to a different spectral typing method; ``k" refers to the star's spectral type based on Ca K absorption; ``h" refers to the star's hydrogen type; ``m" refers to the star's metal-type.} The oft-quoted spectral type for the star of A0 is clearly too hot. Based on the hydrogen type of A6, we adopt an effective temperature of T$_{\rm eff}$ = 8000\,$\pm$\,150 K on the modern dwarf T$_{\rm eff}$ vs. spectral type scale from \cite{pecaut13}. Plotting the star's updated position on an HR diagram (Fig. \ref{fig:age}), along with isochrones from the evolutionary tracks of \cite{bressantracks}, we estimate that HD 32297 is older than \about 15 Myr and younger than \about 0.5 Gyr. The star's kinematics and the debris disk's high fractional luminosity both point to a young age for the star (\about 10-20 Myr). However, we conservatively adopt a stellar age of 100 Myr for this study because no planets are detected, meriting conservative upper limits on planet mass.

We estimate HD 32297's luminosity as log(L/L$_{\odot}$) = 0.79 $\pm$ 0.09 dex. For three sets of \cite{bertelli09} tracks
of varying composition, allowing only for varying T$_{\rm eff}$ and log(L) points that were ``physical" (i.e. not below the zero age main-sequence), and assuming protosolar composition, we determine the mass of HD 32297 to be 1.65 $\pm$ 0.1 M$_{\odot}$. This is much smaller than the oft-quoted higher mass value (\about 3 M$_{\odot}$) that corresponds to a star of spectral type A0.

\subsection{Limits on Planets}
In addition to being sensitive to scattered light from debris disks at \lprime, imaging at this wavelength is also particularly sensitive to self-luminous planets, which are expected to become redder (from 1-5 \microns) as they age and cool \citep{burrows,baraffe}. Understanding the connection between massive planets and debris disks is critical to constraining planetary system architectures and consequently planet formation. 

By inspection of the SNRE map in Fig. \ref{fig:finaldisksnre}, only one source stands at 5$\sigma$ above the noise (the typical minimum threshold for detecting imaged exoplanets). However this feature, located \about 1\asec ~below the star to the southeast, is not point-like in the final reduced image (Fig. \ref{fig:finaldisk}), is not very symmetrical, and does not persist for different reduction methods (e.g., varying the number of modes used in PCA). Therefore we do not treat this as a real astronomical source. Other than this feature, there are no point-source features outside the disk at S/N $>$ 5 in the final \lprime ~image. 

\begin{figure*}[t]
\centering
\subfloat[]{\label{fig:planets1}\includegraphics[scale=0.25]{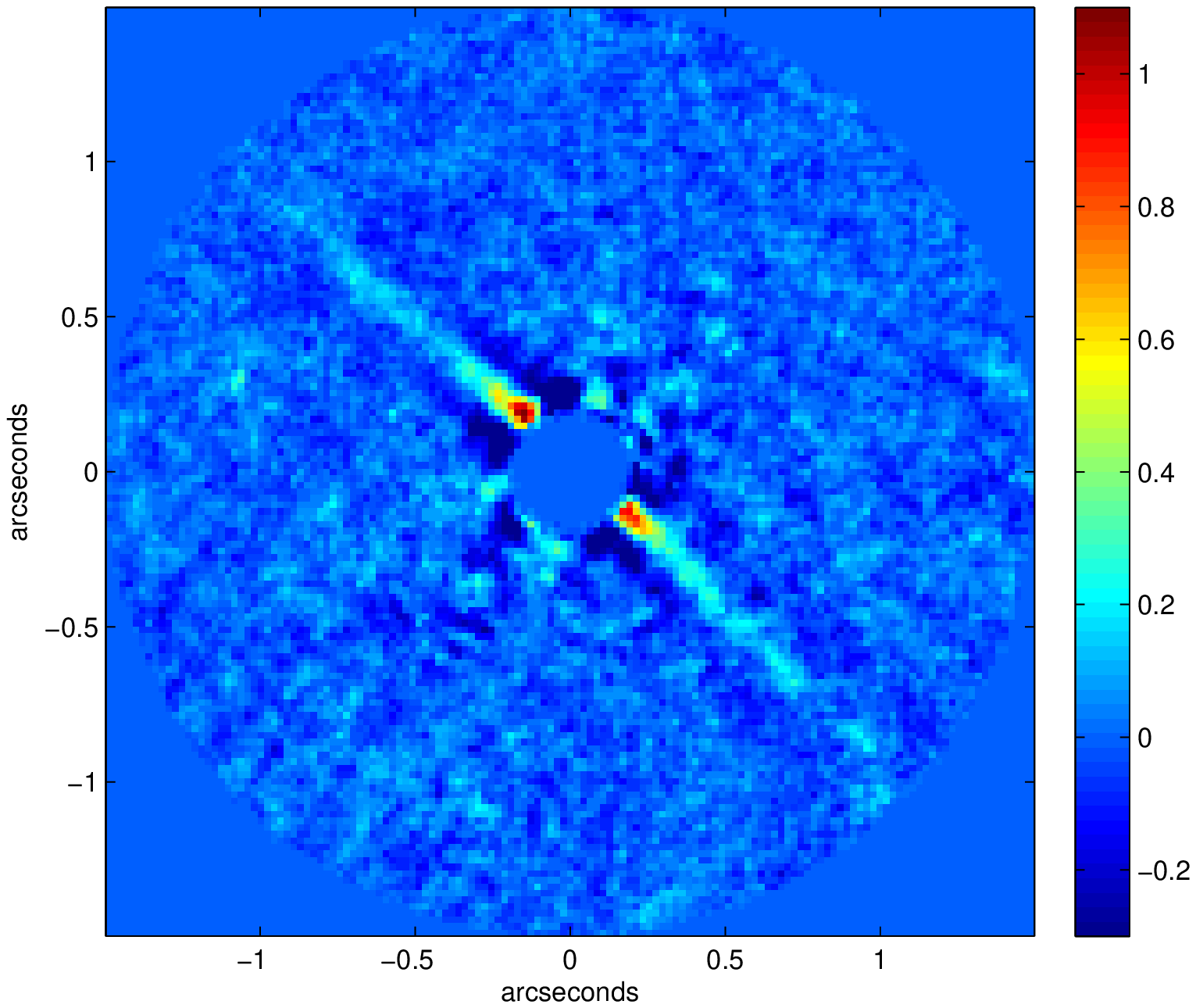}} 
\subfloat[]{\label{fig:planets2}\includegraphics[scale=0.25]{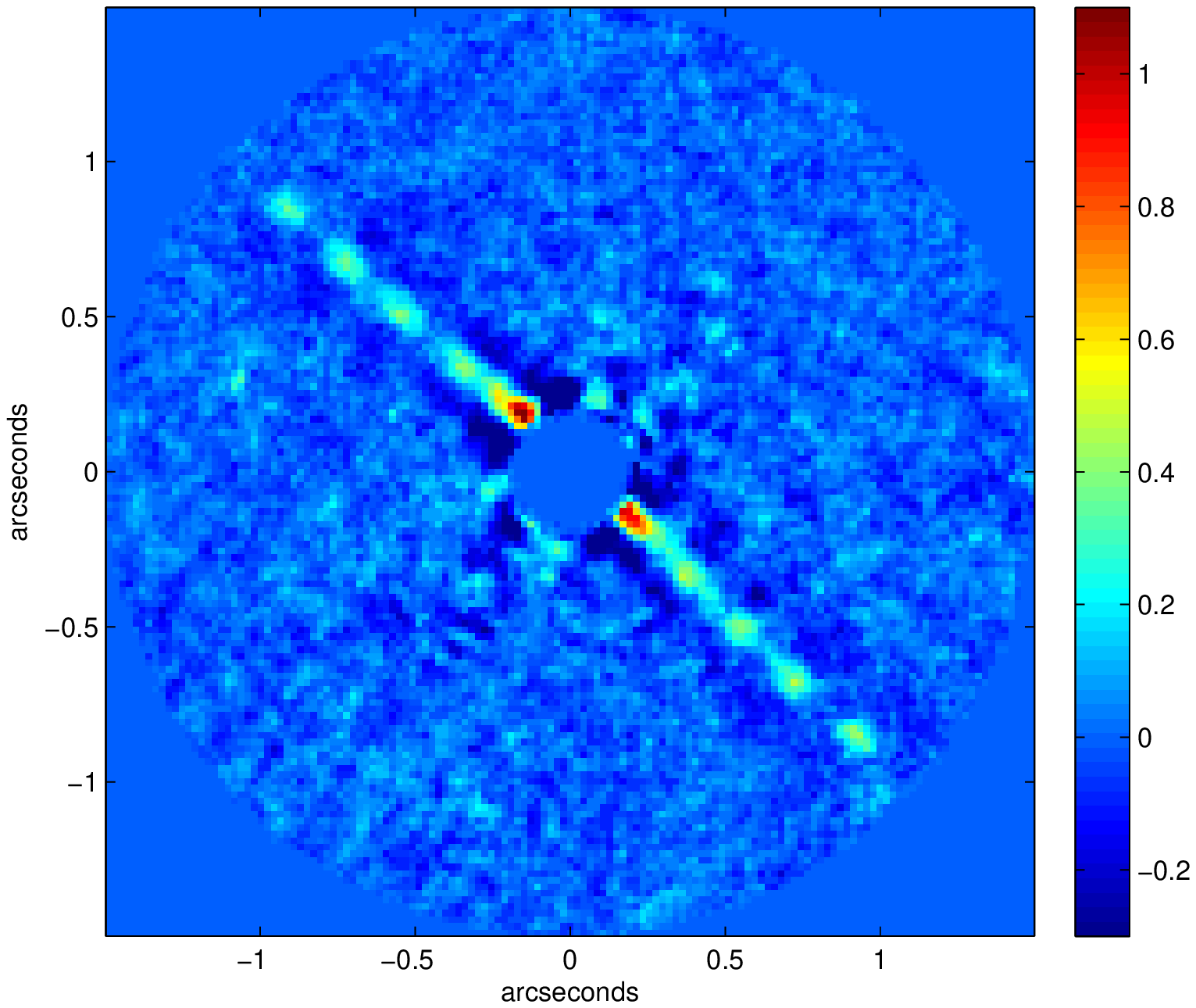}} 
\subfloat[]{\label{fig:planets3}\includegraphics[scale=0.25]{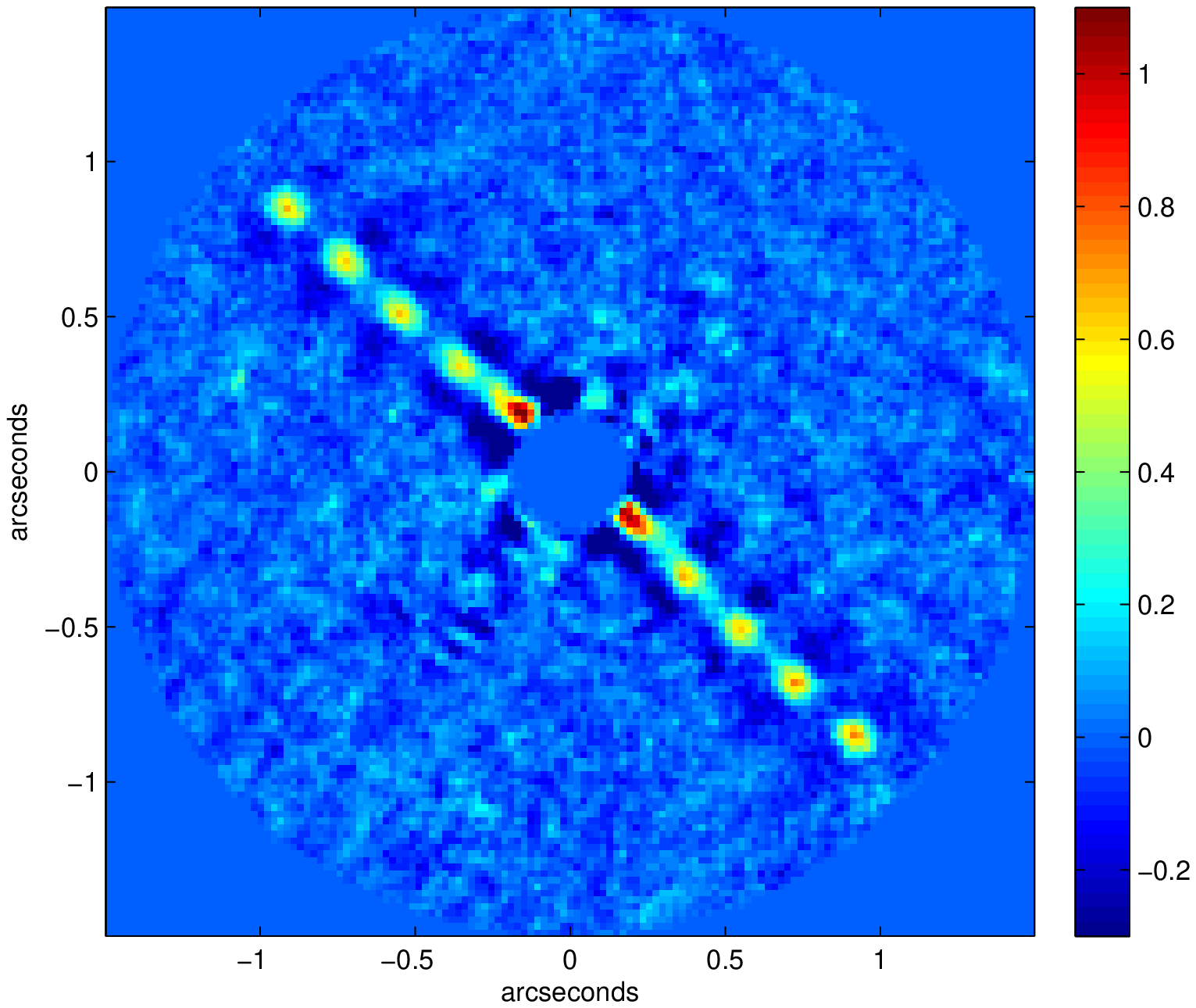}} \\
\subfloat[]{\label{fig:planets4}\includegraphics[scale=0.25]{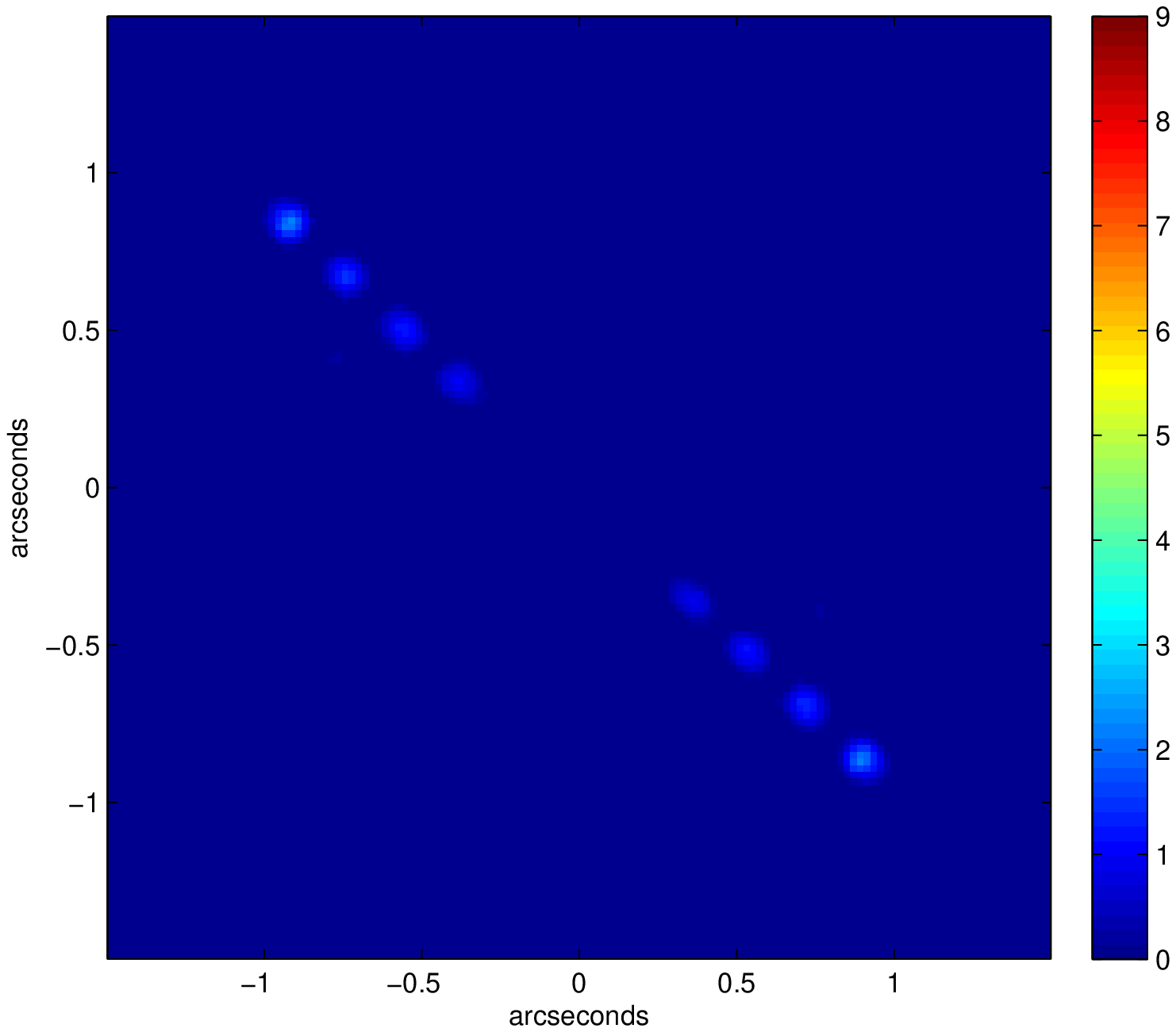}} 
\subfloat[]{\label{fig:planets5}\includegraphics[scale=0.25]{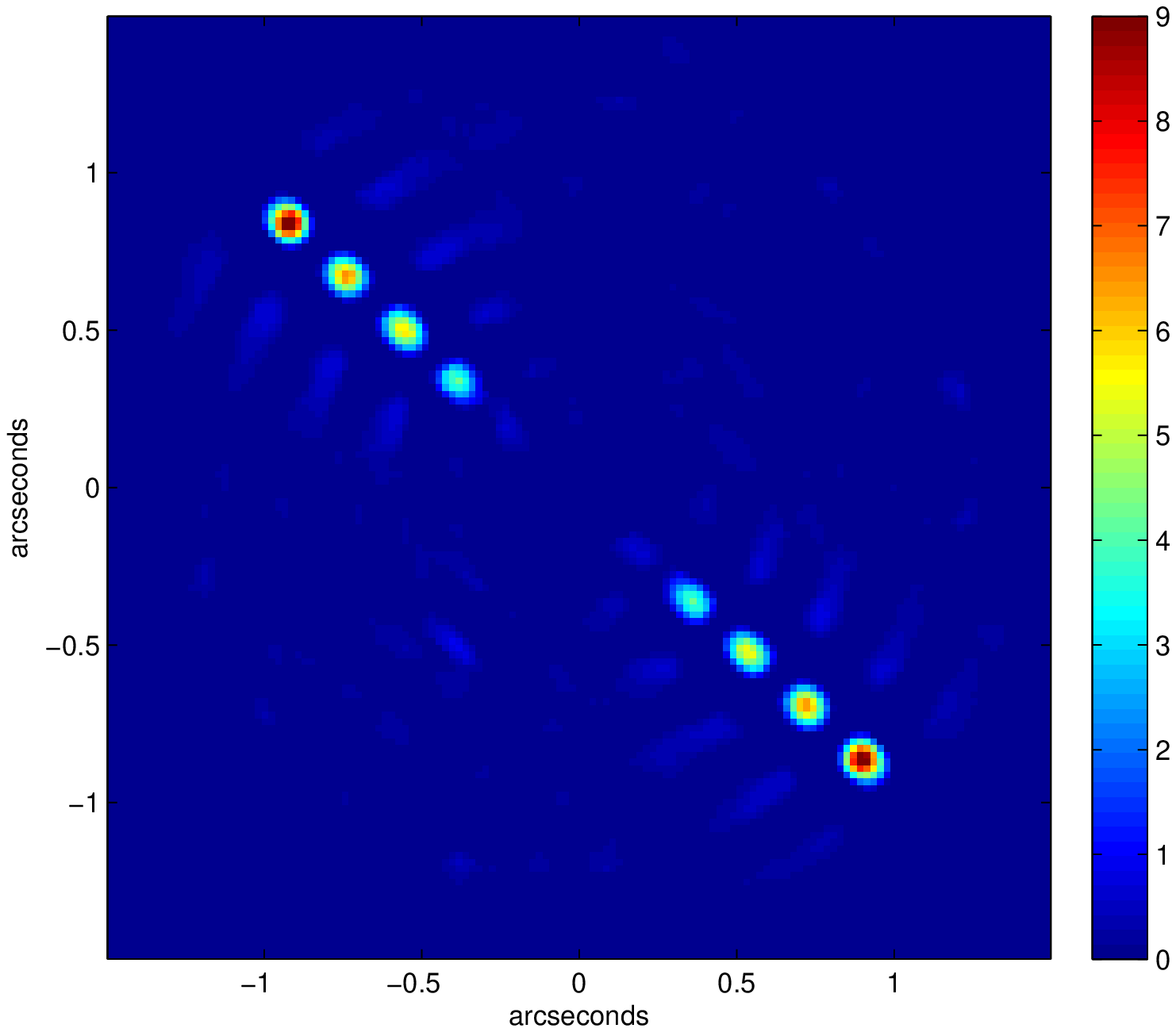}} 
\subfloat[]{\label{fig:planets6}\includegraphics[scale=0.25]{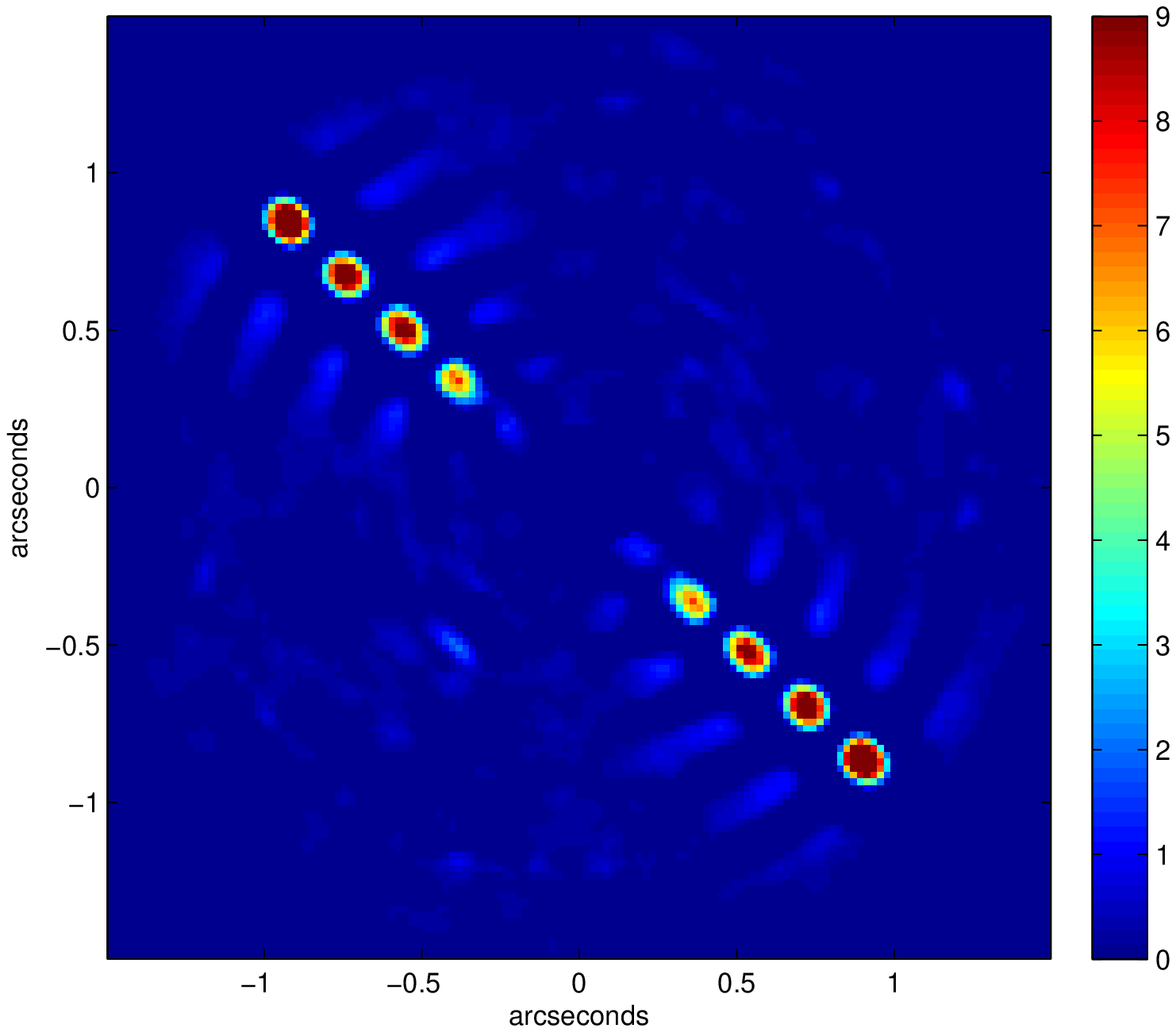}} 
\caption[Artificial planets inserted into the HD 32297 debris disk at varying brightnesses]{Artificial planets of varying brightnesses inserted into the HD 32297 debris disk and recovered. The top row consists of the re-reduced images with the planets inserted, and the bottom row is the corresponding S/N of the detections (after subtracting the S/N of the disk itself). From left to right, the brightnesses of the planets increase, with the leftmost panel showing no 5$\sigma$ detections (planets that are $10^{-5}$ times fainter than the star), the middle panel showing detections beyond 0\fasec 75 (contrast level of $5\times10^{-5}$), and the rightmost showing successful detections for all planets at $\geq$ 0\fasec 5 (contrast level of $5.9\times10^{-5}$).}
\label{fig:planetlimits}
\end{figure*}

To ascertain what planets we could have detected, we assume that any planets in the HD 32297 system must currently reside within the debris disk itself and therefore insert artificial planets into the midplane of the disk. The insertion of artificial sources into the disk is a valid method because the disk is likely to be optically thin, so the signals from any real embedded planets should travel to Earth relatively unimpeded. We vary the brightnesses of the artificial planets and re-reduce the data until each is recovered at $\geq$ 5$\sigma$ confidence. The S/N is calculated as the peak pixel value in the SNRE map at a given position. Artificial planets are made by extracting the central 0\fasec 094 (=FWHM at \lprime) of the unsaturated photometric image of HD 32297. The reduction pipeline for the planets uses the same parameters as were used to detect the disk at high S/N except that we set $K = 5$. 

\begin{figure}[h]
\centering
\includegraphics[scale=0.43]{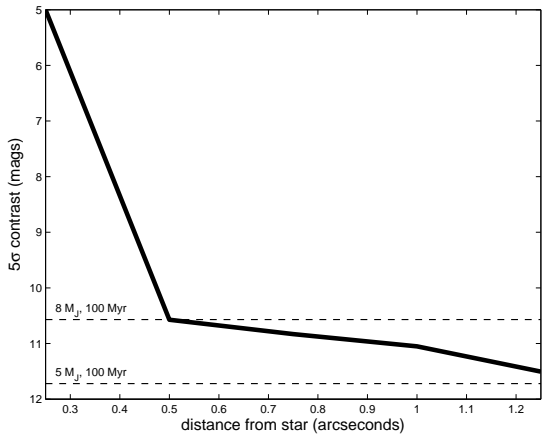} \\
\caption{Limits on the masses of planets that could have been detected at $5\sigma$ confidence in our \lprime ~dataset (solid black line). The dashed lines correspond to the contrast (in mags) 100 Myr old planets would have in this system from \cite{baraffe}. We rule out planets more massive than 8 \mj ~at projected separations $\geq$ 0\fasec 5, and planets more massive than \about 6 \mj ~beyond 1\fasec 25.}
\label{fig:32297masslimits}
\end{figure}

When calculating the S/Ns of the recovered planets, we have to be careful about how we interpret the values since signals from the planets lie \textit{on top} of the signal from the disk. Therefore any measurement of a recovered planet's S/N must first subtract out the S/N of the recovered disk. After doing this, only planets at a contrast level of $2.5\times10^{-5}$ (11.5 mags) were detected at $5\sigma$ confidence beyond 1\fasec 25. At a contrast level of $5\times10^{-5}$, all planets with separations $\geq$ 0\fasec 75 were successfully detected, and at $5.9\times10^{-5}$ all planets at $\geq$ 0\fasec 5 were detected. Within 0\fasec 5, only planets 100 times fainter than the star could be detected at $>5\sigma$ confidence, and even these become highly elongated due to the increased self-subtraction so close to the star. This self-subtraction could in principle be removed by reducing the data more carefully (e.g., including a minimum azimuthal field rotation before subtracting a PSF image), but such an exhaustive optimization of PCA is unnecessary since it would probably not increase contrast levels here by more than a factor of 10. Fig. \ref{fig:planetlimits} shows three example images, along with their S/N maps, of artificial planets inserted into the disk. Fig. \ref{fig:32297masslimits} summarizes all the planet detection results. Adopting a stellar age of 100 Myr and using the hot-start atmospheric models of \cite{baraffe}, we rule out planets more massive than 8 \mj ~at projected separations $\geq$ 0\fasec 5 (56 AU), and planets more massive than \about 6 \mj ~beyond 1.\fasec 25 (140 AU).

\section{Modeling the Debris Disk's Dust}
The high S/N images of HD 32297 debris disk at multiple wavelengths provide a unique window into the dust grain properties within the disk. Under the assumption that a single population of grains can explain all the observations, we set out to test the recent models of the HD 32297 disk from D13, which are constrained by observations of the disk at \ks band \citep{boc32297} along with detailed FIR SED modeling. The primary structure of their modeled debris disk is that of at least one component with a sharp edge at 110 AU and a drop-off in surface density with increasing radius. An interior, warmer component is preferred to fit an additional hot component of dust (D13, \cite{currie32297}), but this is unobservable at the current inner working angles. Any model must be able to reproduce the SB distributions in Fig. \ref{fig:sbplots}.

\subsection{Scattered light model of an optically-thin edge-on disk}
\begin{figure*}[t]
\centering
\subfloat[]{\label{fig:s59}\includegraphics[scale=0.30]{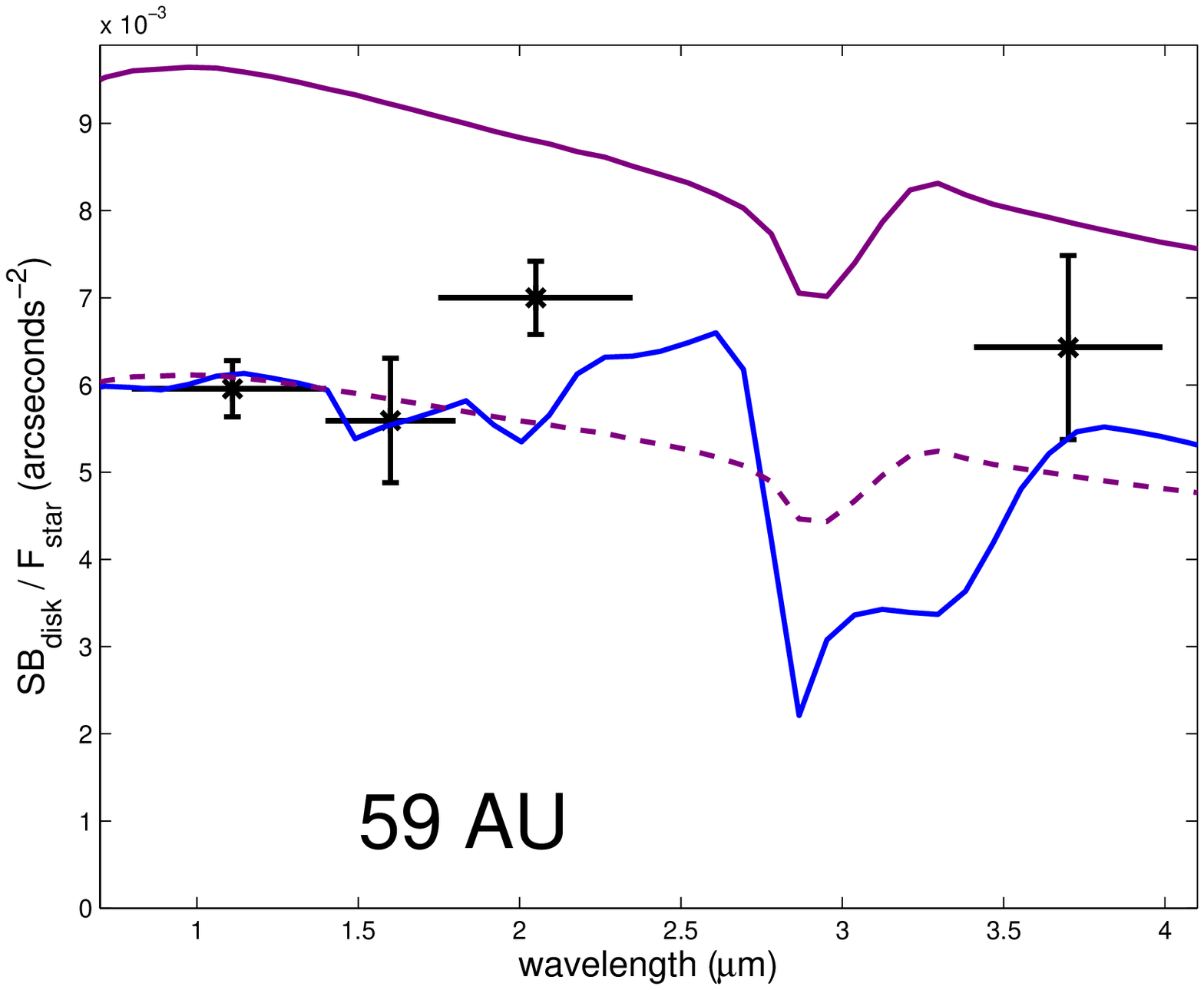}} 
\subfloat[]{\label{fig:s68}\includegraphics[scale=0.30]{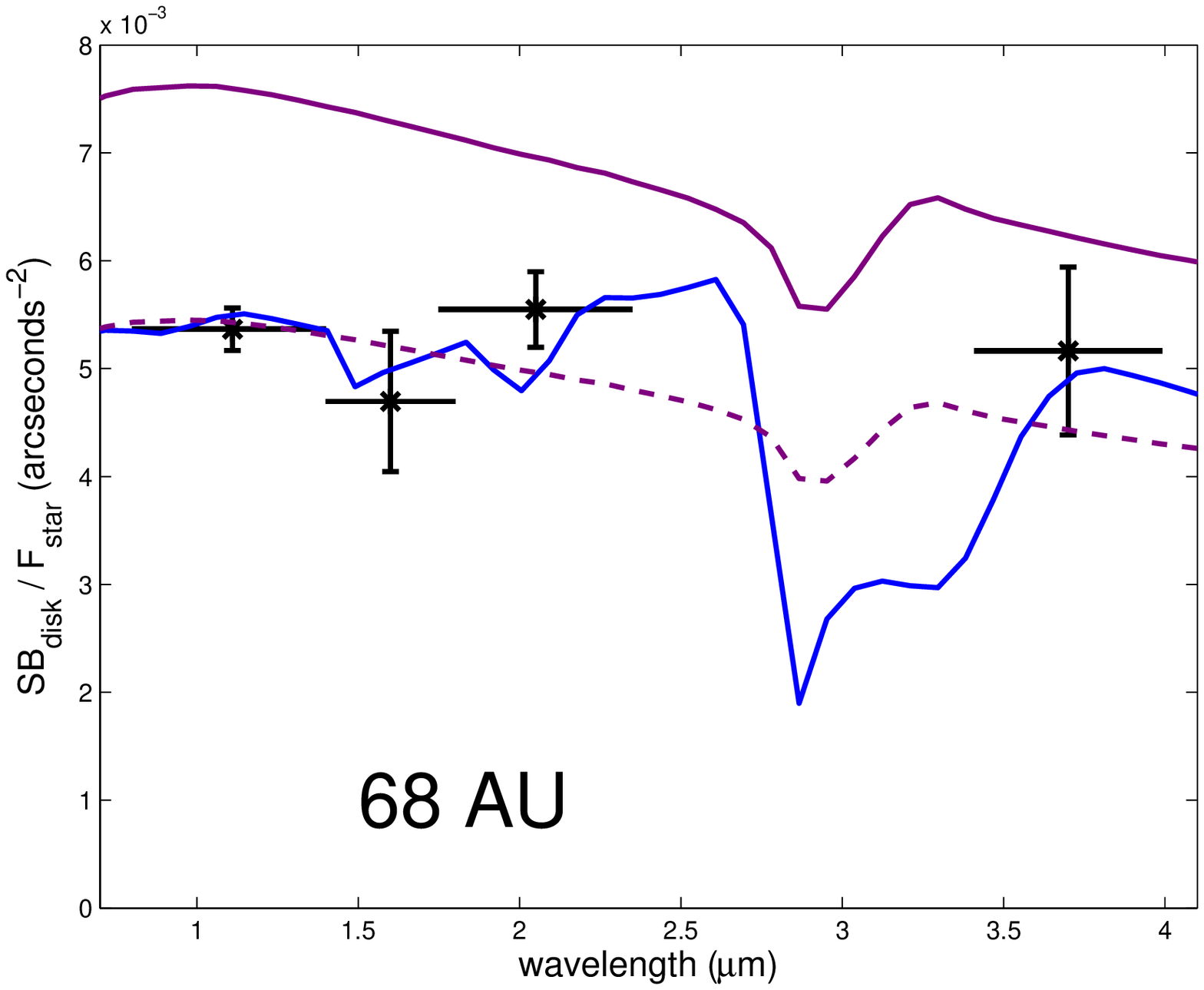}} 
\subfloat[]{\label{fig:s76}\includegraphics[scale=0.30]{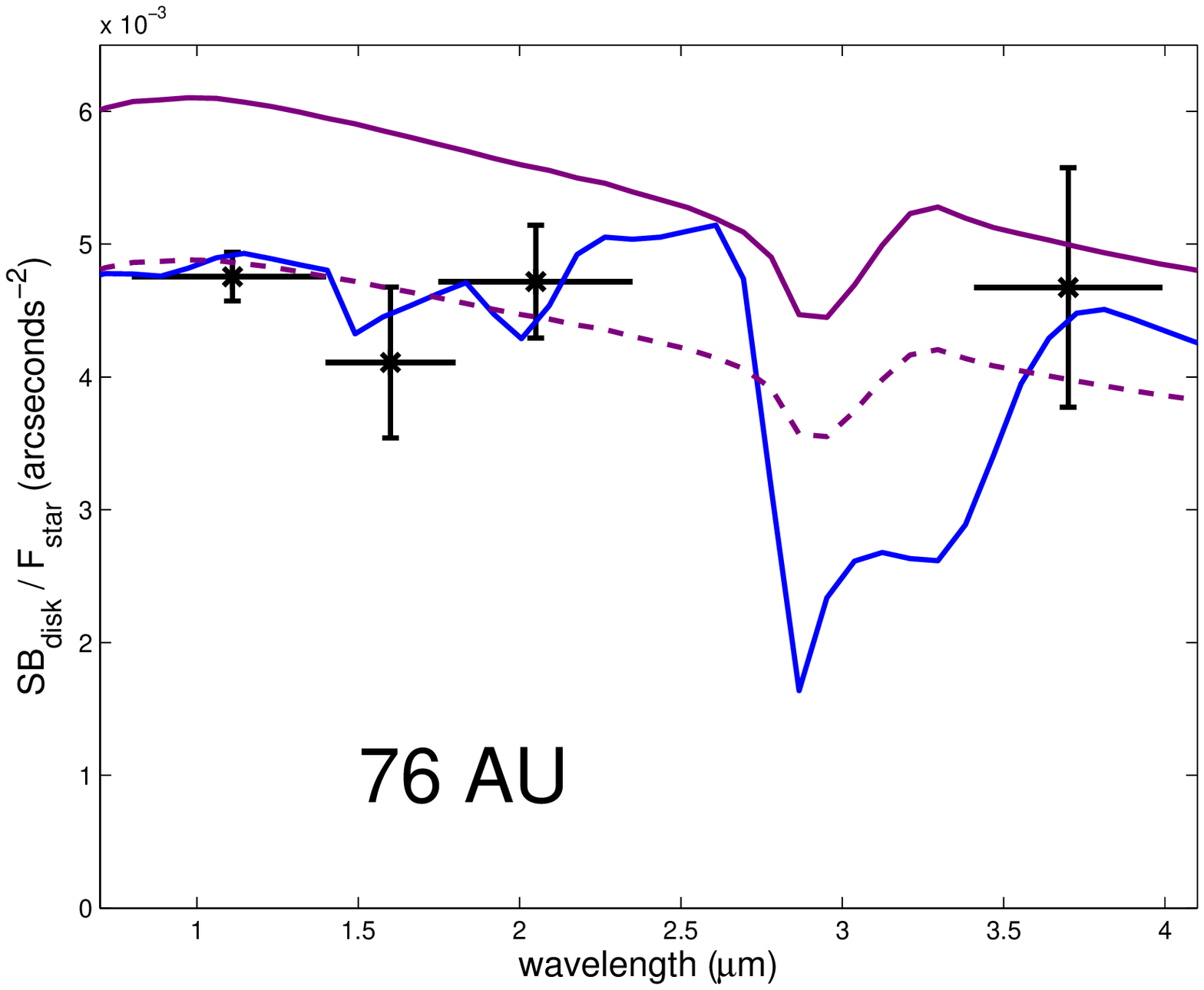}} \\
\subfloat[]{\label{fig:s85}\includegraphics[scale=0.30]{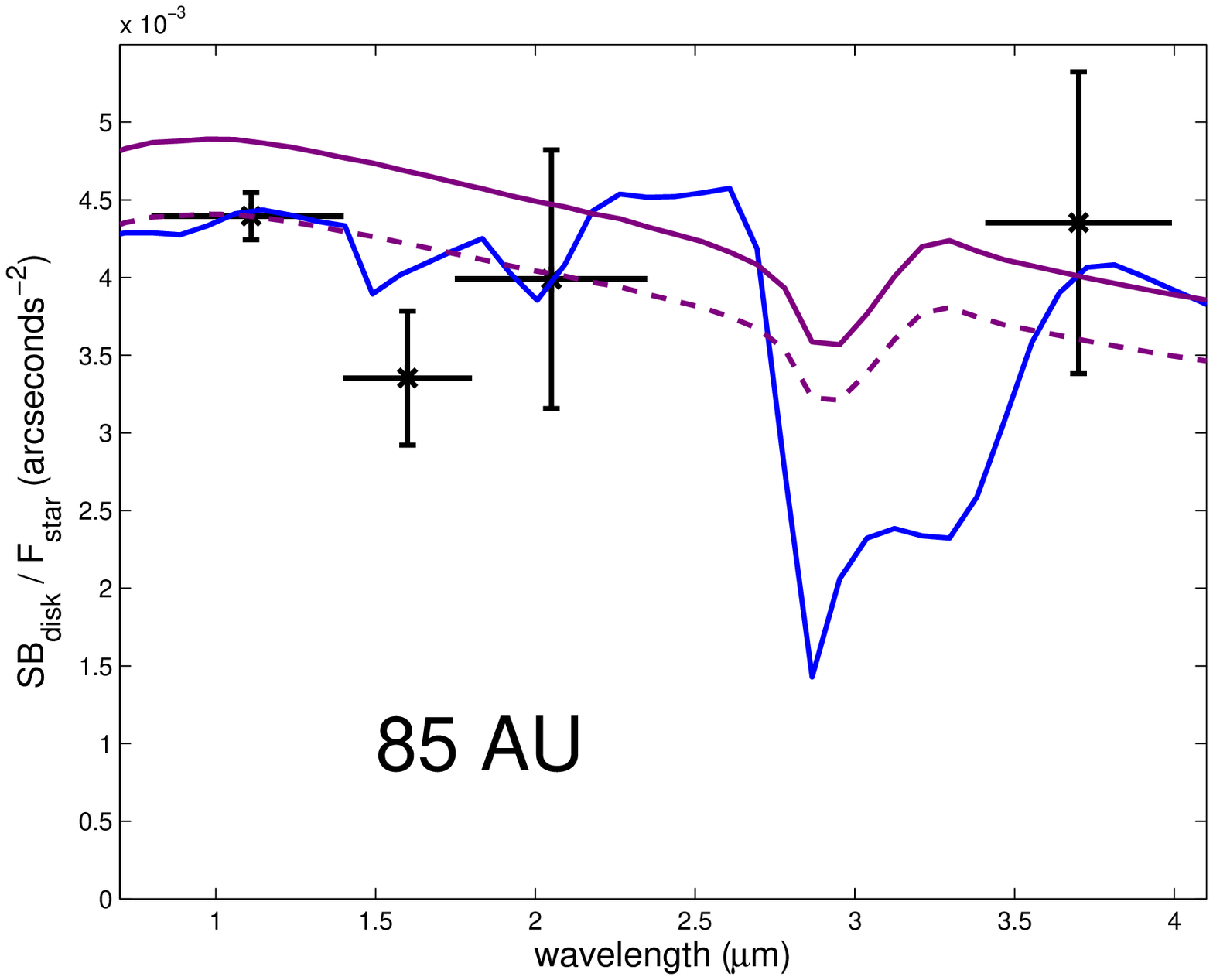}} 
\subfloat[]{\label{fig:s93}\includegraphics[scale=0.30]{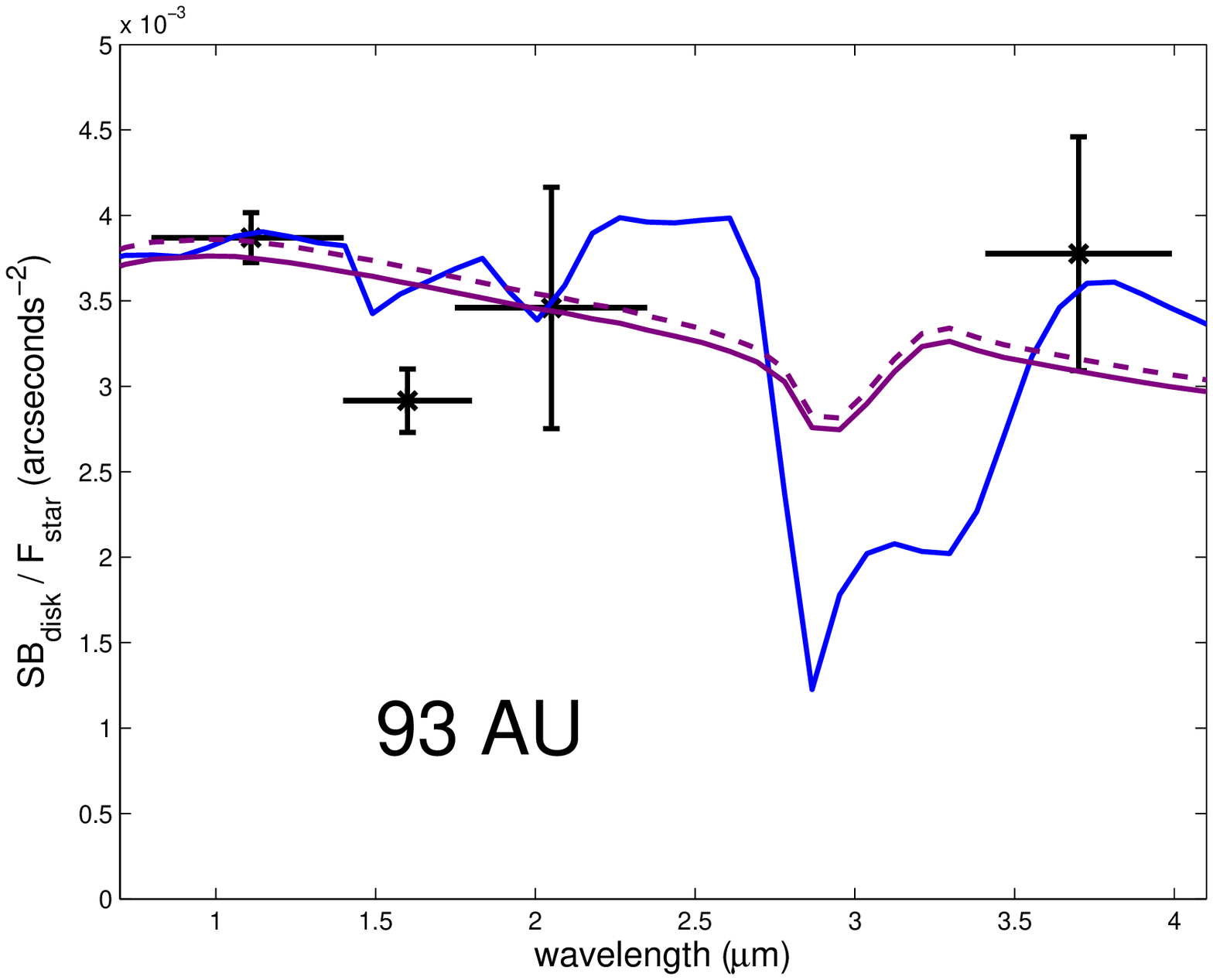}} 
\caption{HD 32297 debris disk spectrum from 59-93 AU, averaged at each distance between the northeast and southwest lobes (black stars), along with the D13 model spectrum (solid purple lines), the modified D13 model (dashed purple lines), and the pure water ice model (solid blue lines). The horizontal black lines denote the filter widths at each wavelength. The original D13 model agrees poorly with the disk's spectrum, with a reduced chi-square of 17.9. The modified D13 model is a better match, with a reduced chi-square of 2.87. The best match to the data is the pure water ice model, with a reduced chi-square of 1.06.}
\label{fig:spectrum}
\end{figure*}

We construct a model of the disk in a similar fashion to \cite{currie32297} (and references therein). An analytical density distribution of dust is generated in a 3 dimensional array and sampled in a Monte Carlo fashion with 2 million particles representing a model dust population. Scattering angles are calculated for the density distribution with a given PA and inclination. Inputs to the model include a cross-sectional averaged asymmetry parameter $<g>$, which can be used to model the forward scattering nature of a grain model in a self-consistent fashion assuming a specific grain size distribution \citep{beust06,wolf2004}:

\begin{equation}
<g> = \frac{\int_{a_{\rm min}}^{a_{\rm max}} \! n(a) C_{\rm sca}(a) g(a) \, {\rm d}a}{\int_{a_{\rm min}}^{a_{\rm max}} \! n(a) C_{\rm sca}(a)\, {\rm d}a},
\end{equation}

where $n$ is the density of the dust in the disk and $C_{\rm sca}$ is the scattering cross-section. The variable $<g>$ can create an approximate phase function as a function of scattering angle $\Phi(\theta)$ for the dust under the assumption of a Henyey-Greenstein (HG) functional form:

\begin{equation}
\Phi (\theta) = \frac{1}{4\pi}\frac{(1-<g>^2)}{(1+<g>^2-2\cos{\theta})^{1.5}}.
\end{equation}

The code can include linear combinations of $<g>$ parameters, which might be appropriate for debris disks and can reproduce the phase function that has been found for the zodiacal dust in the solar system \citep{currie32297,hong85}. To convert an observed disk SB into a mass, one must solve for the combination of both the phase function of the dust and the size-averaged cross section of the dust $<C_{\rm sca}>$:

\begin{equation}
<C_{\rm sca}> = \int_{a_{\rm min}}^{a_{\rm max}} \! n(a) C_{\rm sca}(a) \, {\rm d}a.
\end{equation}

We calculated scattering cross-sections $<C_{\rm sca}>$ and $<g>$ using the real and imaginary parts of the complex indices of refraction for the best-fitting grain model (provided kindly by J. Donaldson, private communication) from the code $miex$, which has been designed specifically for fast modeling of debris disks with a size distribution of dust \citep{wolf2004,ertel11}. The grain model used here (and by D13) is a 1:2:3 mixture of 90$\%$ porous silicates, carbon, and water ice grains 2.1-1000 \microns ~in size; such compositions may also be appropriate for other debris disks (e.g., \citealt{hd181327ice}). Models were generated for each image (wavelength) with the appropriate pixel scale and sampled in a similar fashion to our SB profiles (Section \ref{sec:sbmeasures}). A scaling factor for each lobe of the disk was calculated by ratioing the models with the observed disk SB profiles as a function of wavelength. For the disk's density distribution, we used the best-fitting distribution in \cite{boc32297}, based on analysis of their \ks band disk data, because D13 also used this model for their geometrical constraints:

\begin{equation}
n(r) = n_{0} \sqrt{2} \left(\left(\frac{r}{110 {\rm AU}}\right)^{-2\alpha_{out}} + \left(\frac{r}{110 {\rm AU}}\right)^{-2\alpha_{in}} \right)^{-1/2},\
\end{equation}
where $n_{0}$ is the midplane number density at the reference distance of 110 AU, and $\alpha_{out} = -5$ and $\alpha_{in} = 2$. We also kept their assumption of $<g>$ = 0.5, which is somewhat degenerate with the choice of an interior steep power law drop-off in density interior to 110 AU. It is also not consistent with Mie calculations of the expected $<g>$, which is closer to 0.99.


\subsection{Comparison to original D13 cometary grains model}
For the D13 model to be accepted as a good match to the data, it must reproduce the disk's SB at all wavelengths and distances from the star. This can be measured by calculating the reduced chi-square. We accomplish this by summing up the squared difference between the model disk SB and the real disk SB at all wavelengths and at all disk locations, divided by the number of degrees of freedom and the uncertainties squared. We measured this value to be 17.9, indicating a poor fit. To illustrate, Fig. \ref{fig:spectrum} shows the spectrum of the disk at 59-93 AU (black stars), along with the D13 cometary grains model (solid purple lines). The SB values for the northeast and southwest lobes of the disk have been averaged at each wavelength because the D13 model is axisymmetric. 

In general, this model overpredicts the disk's SB and is therefore a poor overall fit. However, at \about 85-93 AU (Fig. \ref{fig:s85} and Fig. \ref{fig:s93}), the model does a much better job of fitting the data. This makes sense since \cite{boc32297} computed the model disk from images in which the disk was only detected beyond 0\fasec 7 (80 AU). More importantly, the discrepancy between the model fit at small and large separations demonstrates the need for resolved imaging with small inner working angles (such as our \lprime ~image). 

\subsection{Comparison to modified D13 cometary grains model}
Seeking to obtain an alternate model fit to the data, we modified the density distribution of the original D13 cometary grains model. This is necessary because the disk's actual SB at 1-2 \microns ~is shallower (lower SB power-laws) than the model disk's SB. Therefore we modified the model's dust distribution to better reproduce the data. Specifically we tested different values of $\alpha_{out}$ and $\alpha_{in}$, picking $\alpha_{out} = -2$ and $\alpha_{in} = 5$ based on chi-square minimization. This modification resulted in a much improved reduced chi-square value of 2.87. This model is shown in Fig. \ref{fig:spectrum} as the dashed purple lines. 

\subsection{Comparison to pure water ice model}
\begin{figure}[h]
\centering
\includegraphics[scale=0.45]{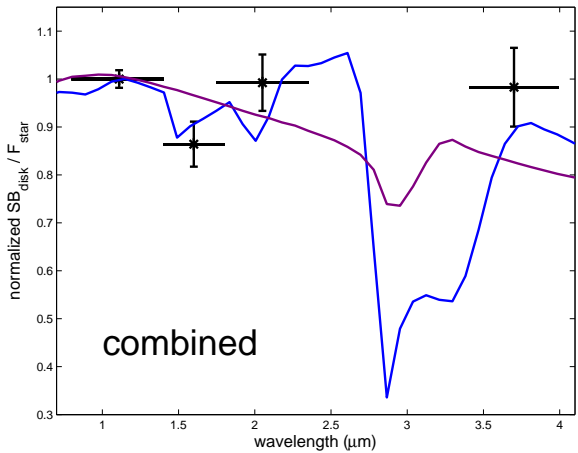}
\caption{Combined spectrum of the disk from 59-93 AU, obtained by normalizing the spectra from Fig. \ref{fig:spectrum} at 1.1 \microns ~and then taking the median value at each wavelength (black stars). The horizontal black lines denote the filter widths at each wavelength. The solid purple line corresponds to the D13 model spectrum, adjusted such that the residuals between the data and the model are at a minimum. The solid blue line corresponds to the pure water ice model spectrum, adjusted in the same fashion. After scaling the D13 model spectrum, the pure water ice model has a chi-square value \about 3$\times$ lower, indicating a better spectral fit to the data, but additional data at 3.1 \microns ~is required to help distinguish between comets and pure water ice.}
\label{fig:finalspectrum}
\end{figure}

We also tested a pure water ice model, since the inclusion of water ice significantly improved the model fits to the FIR SED of the disk (D13). Simpler compositions like astronomical silicates are not modeled because these produced poor fits to the FIR SED data (D13), and more complex models like tholins \citep{4796organics} resulted in poor fits to the data . The pure water ice model disk had the same density distribution as the modified D13 model ($\alpha_{out} = -2$ and $\alpha_{in} = 5$), except the dust grains were hard spheres 3 \microns ~in size. This model is the best fit to all of the available NIR data (Fig. \ref{fig:spectrum}; reduced chi-square = 1.09).

If we ignore the requirement that the models must match the disk's SB distribution, allowing the model spectra to be arbitrarily scaled to match the disk's spectrum, we can determine which model is the best \textit{spectral} fit to the data. To increase the S/N of the data, we normalized the disk spectra from 59-93 AU at 1.1 \microns ~and computed the median value at each wavelength. These values are shown in Fig. \ref{fig:finalspectrum} (black stars). We then scaled the original D13 model spectrum until the residuals between the data and model were minimized. This is shown as the solid purple line in Fig. \ref{fig:finalspectrum}. After optimizing the spectral fit, the D13 model is blue and underpredicts the disk's SB at 3.8 \microns, while the real disk's spectrum is gray within the uncertainties. We repeated this procedure for the pure water ice model. The chi-square value for the water ice model is \about 3$\times$ lower than the D13 model, indicating this model is still preferred over the cometary grains model based on the available data.

\section{Discussion}
\subsection{Disk Structure}
Fig. \ref{fig:sbplots} and Fig. \ref{fig:PAs} allow us to characterize the structure of HD 32297's debris disk. The SB asymmetry between the two sides of the disk from 0\fasec 5-0\fasec 8 (56-90 AU) at \lprime ~(and \ks band from \cite{currie32297}) suggests that the northeastern lobe of the disk has a deficit of dust grains here. Furthermore, by inspection of the SB profile power-law indices in Table \ref{tab:indices}, we can see that interior to 0\fasec 5 (56 AU), the disk's SB rises very sharply, whereas outside this location the SB increases more slowly. This is exactly the behavior one would expect to see for a disk with a second interior disk component located close to the star. A single-component disk would have a SB profile that either steadily increased close to the star with the same power-law index as further from the star, or a SB profile that flattened out close to the star (e.g., HD 15115 as in \citealt{mehd15115}). The steep rise in SB close the star was also seen at \ks band \citep{currie32297}, corroborating this feature. The inner component could be located at $\lesssim$ 50 AU but needs to be confirmed with additional imaging that is sensitive to scattered light close to the star. 

This is not the first time a second inner disk component has been suggested for the HD 32297 debris disk. D13 favored a warm inner component for their modeling, and \cite{hd32297fitz} used 11 \microns ~images to suggest a population of dust grains that peaks near \about 60 AU. Such thermally-emitting grains may also be contributing to the scattered light that has been detected at 2-4 \microns. \cite{currie32297} also preferred multiple belts, including a component at \about 45 AU, to explain the disk's observed SB and SED. Our \lprime ~data appear to support the notion of a second interior belt, but additional data with very small inner working angles would help validate this explanation.

The midplane offset profiles show that the disk is bowed close to the star, agreeing with similar findings at \ks band \citep{currie32297,boc32297,esposito32297}. We do not explicitly model this phenomenon because we have already showed that it can be explained (for both HD 15115 and HD 32297) by a highly-inclined ring-like disk consisting of forward-scattering grains \citep{mehd15115,currie32297}. 

No high-mass (8 \mj) planets at projected separations $\gtrsim$ 56 AU currently reside in the disk based on our \lprime ~imaging. If such planets exist in this system, their projected separations must be very small, implying either small semimajor axes or an unlucky epoch of imaging. The null detection of high-mass planets in this system, though disappointing, should not be surprising, since there is now copious evidence that such planets are rare at large separations from their stars (e.g., \citealt{wahhaj}). While lower-mass planets may still reside in the HD 32297 system, detecting them will require more advanced imaging capabilities than are currently available. This is further complicated by the likelihood of such planets residing in the midplane of the disk, which if bright enough, can hide planets. Residuals from PSF subtraction (especially in ADI datasets) can themselves also resemble point sources, and since they can be found anywhere in an image (Fig. \ref{fig:finaldisksnre}), including in the midplane of the disk, distinguishing between real and artificial planets is difficult for systems like HD 32297. Future imaging searches for planets around this star must adequately take into account the effects of both the disk itself and the PSF residuals.

\subsection{Dust Grain Composition}
D13 found that comet-like porous grains consisting of silicates, carbon, and water ice were the best match to the HD 32297 disk SED at wavelengths longer than 25 \microns. Based on our analysis of the disk's 1-4 \microns ~SB and spectrum, we cannot lend further evidence to support this claim, at least with respect to their original model. The overall agreement between the data and the D13 model is poor (reduced chi-square = 17.9). 

This does not necessarily exclude comet-like materials from being present in the disk. We were able to achieve a much better fit simply by altering the dust density distribution in the original cometary grains model. It may be that additional tweaking of model parameters will yield an even better fit. Furthermore D13 reported only one ``comet" combination; other combinations with differing ratios of carbon/silicates/water ice and perhaps the dust's porosity might better match the disk's spectrum at $\lambda > 1$ \microns. Because the spectral coverage is currently sparse from 1-4 \microns, complex cometary grain modeling is beyond the scope of this paper. 

The best fit to the data is achieved by a pure water ice model, likely due to the fact that the model is gray from 1-4 \microns, like the disk. This model has a very deep absorption feature near 3.1 \microns, evident in Fig. \ref{fig:finalspectrum}. The feature is also evident in the cometary grains model, though it is much shallower. Therefore a key additional test of the disk's dust grain composition would be obtaining very high S/N photometry of the disk using a narrowband filter centered around 3.1 \microns. This would help determine if the disk's dust contains any water ice (e.g., \citealt{honda}), which could then be used to refine the D13 cometary grains model.

Other dust grain compositions not tested in this work may also better support the available data. For example, a perfectly flat/gray model spectrum might better fit the disk's NIR spectrum, even if such a dust composition is not easily explained. This is why obtaining a disk detection at 3.1 \microns ~is crucial; it can help distinguish between various dust models that might fit the available NIR data.

\subsection{Limitations}
While our pure water ice model is the best match to the scattered light data, it may not be for the FIR data (used by D13). Our model must reproduce all of the available photometry at all stellocentric distances to be accepted as valid; therefore our scattered light modeling should not be considered final. This is especially true because we currently lack data near 3.1 \microns, where our preferred model has a very deep absorption feature. We stress that the goal of this study was to test the originally-proposed D13 cometary grains model; if that failed to fit the data, the goal was to find a reasonable alternative. Pure water ice is one such alternative composition, and it makes predictions that are testable with future data, which can then help refine the original cometary grains model. 

We also note that producing scattered light models of debris disks comes with several problems. The scattering asymmetry parameter, $<g>$, is not self-consistent when computed using the Mie formalism. Additionally the commonly assumed single component HG formalism may not be appropriate for this disk \citep{currie32297}. This would immediately make modeling the dust composition more complicated and beyond the scope of this initial effort. Finally the increasing evidence that the disk may have an inner component at $\lesssim$ 50 AU further complicates the modeling process, since we would then have to consider the possibility of two separate dust populations with two different dust compositions. At this time, it is unclear what the underlying distribution of dust is around HD 32297. Lacking this knowledge, assuming a single disk component as we have in this study is a reasonable starting point.

\section{Summary}
We have presented an imaging detection of the HD 32297 debris disk at \lprime. The disk is detected at high S/N from \about 0\fasec 3-1\fasec 1 (30-120 AU). Based on our \lprime ~imaging, we show that the system does not contain any planets more massive than \about 8 \mj ~beyond 0\fasec 5. 

The disk at \lprime ~is bowed, as was seen at \ks band. This likely indicates that the disk is inclined by a few degrees from edge-on and contains highly forward-scattering grains \citep{mehd15115}. The SB at \lprime ~interior to 50 AU rises sharply, as was also seen at \ks band \citep{currie32297}. This evidence together suggests that the disk may contain a second inner component located at $<$ 50 AU.

Comparing the disk's color at 1-4 \microns ~over the outer portion of the disk ($> 50$ AU) with the recently proposed cometary dust grain model of D13 shows that this model is a poor overall fit to the disk's SB distribution and spectrum (reduced chi-square = 17.9). A modified version of this model produces a much better overall fit to all the data (reduced chi-square = 2.87). The best fit to the data is achieved with a pure water ice model, though this model is not the only possible dust composition. Additional imaging of the disk near 3.1 \microns ~can help constrain the fractional amount of water ice in the dust. This will then help determine how similar the HD 32297 dust grains are to comet-like grains.

\acknowledgments
We thank the anonymous referee for helpful comments that improved this paper. We thank Jessica Donaldson for sharing her disk model and for helpful discussions. We thank the LBT observatory staff for their help operating and maintaining the telescope and its powerful instruments. We acknowledge support for LMIRcam from the National Science Foundation under grant NSF AST-0705296. T.J.R. acknowledges support from the NASA Earth and Space Science Fellowship (NESSF) during his time at the University of Arizona. VB is funded by the NSF Graduate Research Fellowship Program (DGE-1143853). 

\appendix
\section{Surface Brightness Corrections}
\label{sec:corrections}
Determining the SB of the disk as a function of distance from the star at multiple wavelengths requires several steps to ensure the correct quantities are measured \citep{debes}. The true SB of an edge-on debris disk a distance $r$ away from the star is computed as follows:
\begin{eqnarray}
\hbox{SB}_{true}(r) = \hbox{SB}_{measured}(r) \times (\hbox{PSF convolution correction}) \nonumber \\
\times (\hbox{aperture size correction}) \times (\hbox{reduction bias correction}), \nonumber \\
\label{eqn:corrections}
\end{eqnarray}
where the PSF convolution correction (often referred to as ``aperture correction") is employed to account for non-infinite photometric apertures, and the latter two corrections are only necessary if different aperture sizes are used at different wavelengths and if the data reduction pipeline alters the SB of the disk as a function of distance from the star.

We computed the appropriate corrections necessary for the HST/NICMOS and LBTI/LMIRcam \lprime ~data. The PSF convolution correction was computed as follows. We produced unconvolved and convolved model disk images (with parameters determined from the best-fitting model used in \citealt{boc32297}). In total, we generated one unconvolved model image for the HST/NICMOS data (since all the HST data was taken at the same plate scale), and one unconvolved model image for the higher-resolution LBTI/LMIRcam data. Four convolved images (one for each wavelength) were produced by convolving the corresponding unconvolved model image with the appropriate PSF template (either HST/NICMOS or LBTI/LMIRcam). 

We measured the SB in all the HST/NICMOS model images using the same 3 pixel (0\fasec 2262) by 3 pixel boxes as were used on the real disk images. We first computed the SB in each box for the unconvolved images, then we computed the SB values in the convolved images and determined the appropriate correction factors needed to obtain the same SB in the unconvolved and convolved images as a function of distance from the star. 

\begin{figure*}[h]
\centering
\subfloat[]{\label{fig:fakedisks}\includegraphics[scale=0.3]{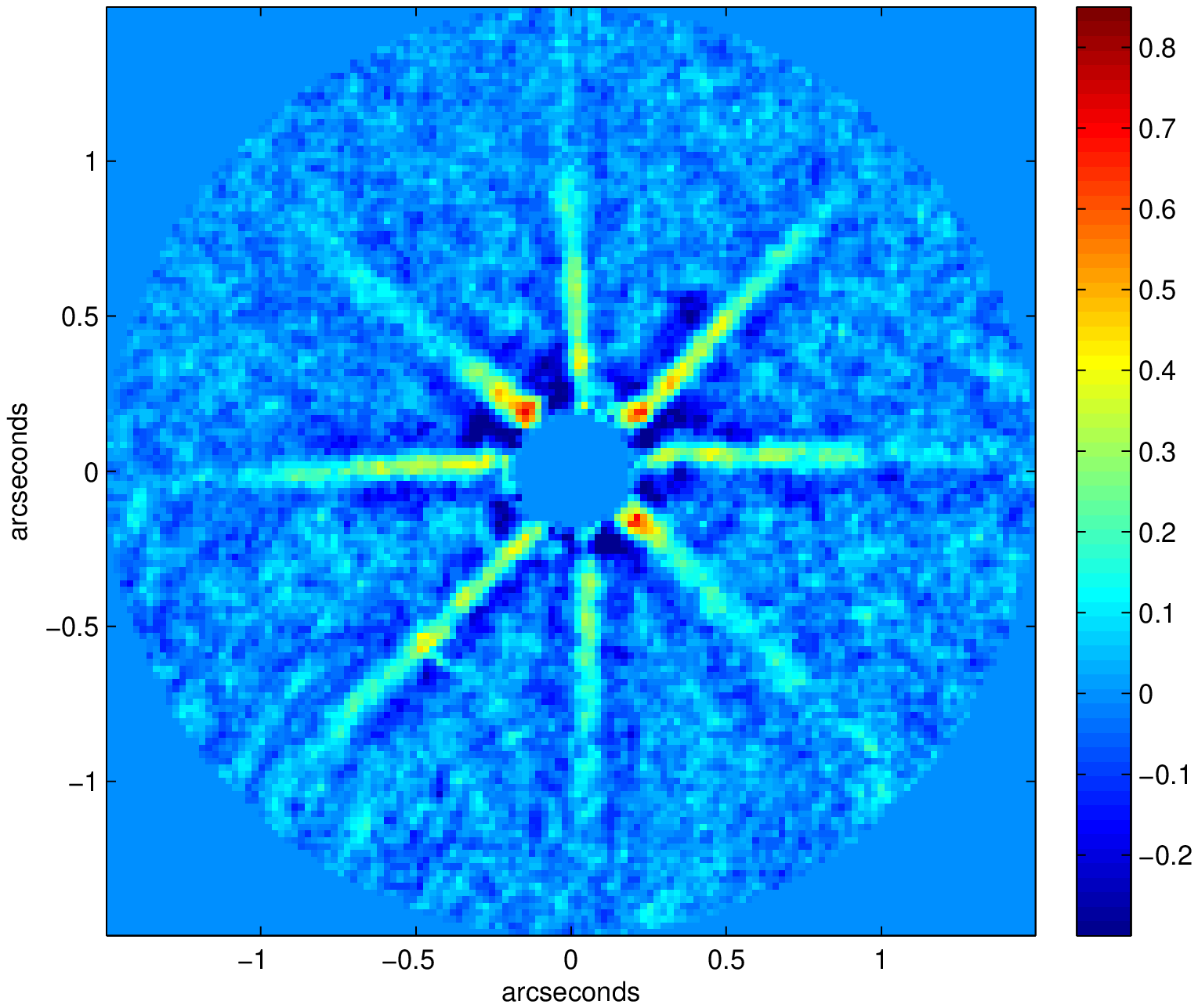}}
 \subfloat[]{\label{fig:pcabias}\includegraphics[scale=0.3]{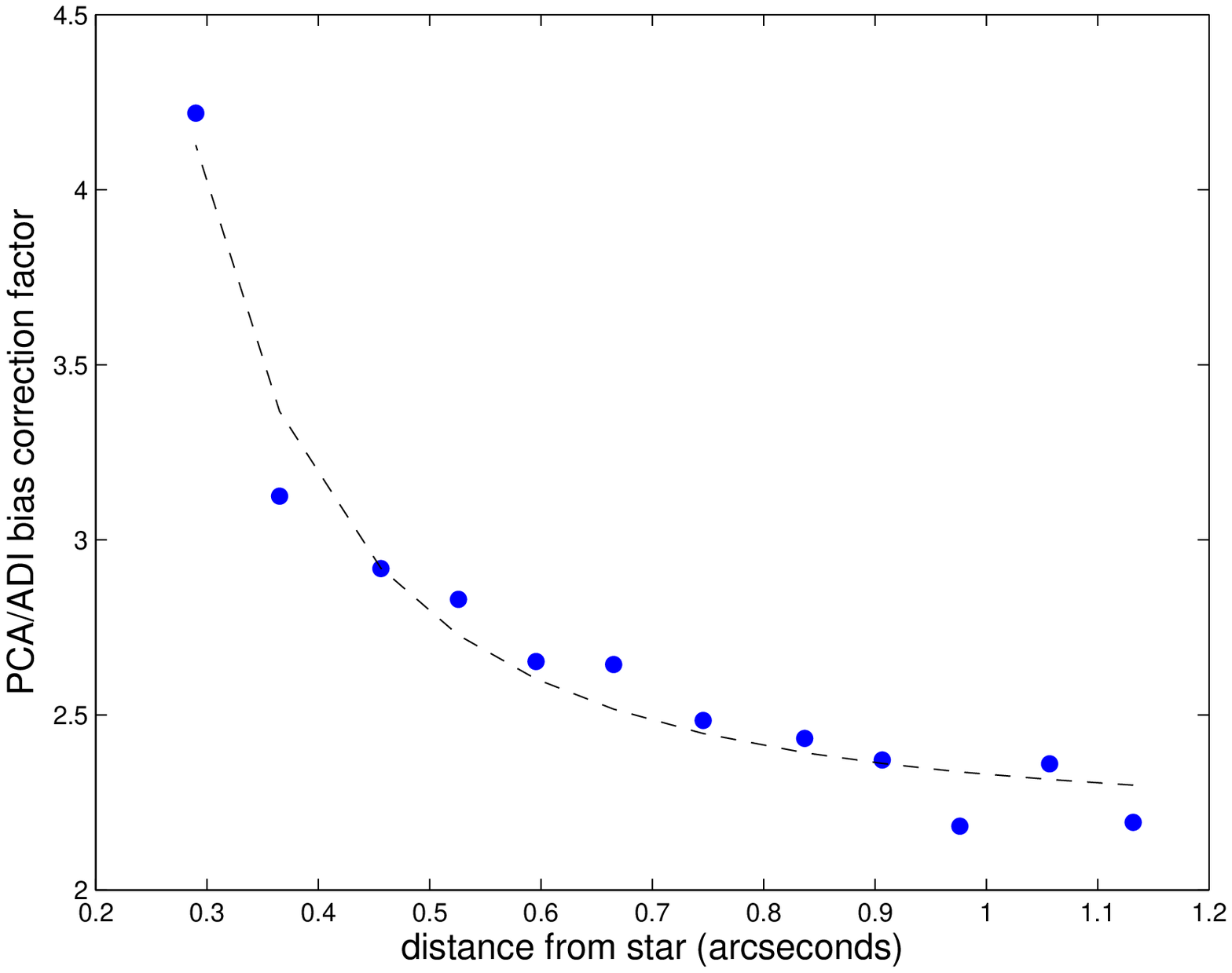}}
 \subfloat[]{\label{fig:otherbias}\includegraphics[scale=0.3]{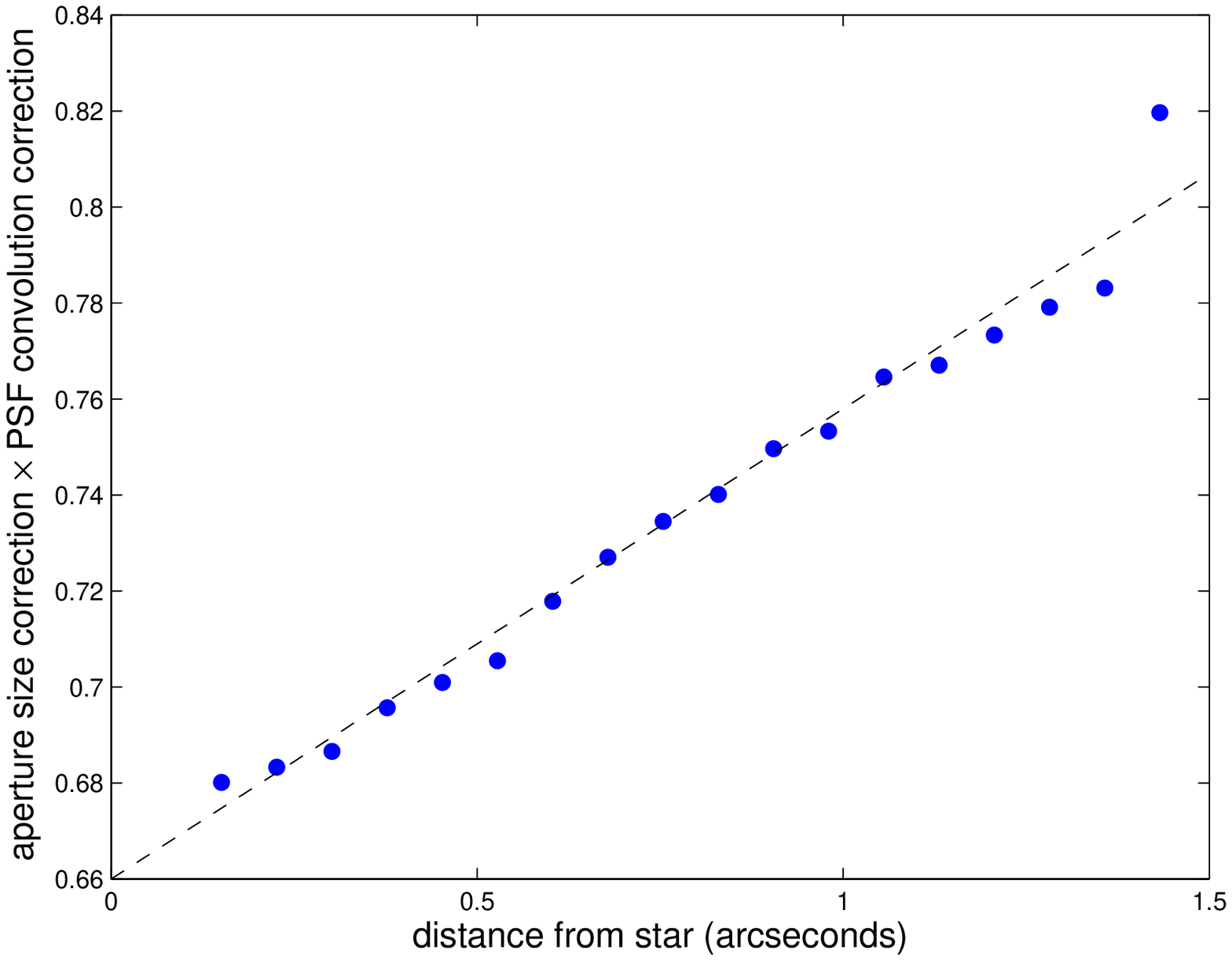}}
\caption{\textit{Left}: Final reduced \lprime ~image of the HD 32297 debris disk along with three artificial disks, all in units of detector counts/s, with North-up, East-left. A 0\fasec 2 radius mask has been added in post-processing. The real disk is located at its nominal PA of \about 47\degrees. All three model disks are easily recovered by the PCA pipeline, allowing correction factors for the real disk's surface brightness to be calculated. \textit{Middle}: PCA/ADI reduction bias correction factors as a function of distance from the star, with a fitted function (dashed line). \textit{Right}: the same, except for the aperture size correction factor $\times$ the PSF convolution correction factor. The correction changes with separation from the star because a non-uniform structure that changes with distance is being convolved with a PSF \citep{debes}.}
\end{figure*}

The aperture size correction was only needed for the \lprime ~data, since a smaller photometric aperture was used. We compared the median SB in the unconvolved model HST image (computed using the 3 pixel (0\fasec 2262) by 3 pixel boxes) with the equivalent quantity in the unconvolved model \lprime ~image (computed using the 5 pixel (0\fasec 106) by 5 pixel boxes). The correction factors were computed as the ratio of these two values as a function of distance from the star. To illustrate, Fig. \ref{fig:otherbias} shows the aperture size corrections $\times$ the PSF convolution corrections for the \lprime ~data.

Finally, the \lprime ~data required a correction to account for the biases inherent in PCA + ADI data reduction (e.g., \citealt{mehd15115,currie32297}). We measured these biases by inserting artificial disks into the raw images at differing PAs, re-reducing the data, and computing the correction factors based on how the SB of the artificial disks changed.\footnote{\cite{esposito32297} showed that forward-modeling was a potentially more accurate method for determining the true SB of edge-on disks. However their method currently requires the use of LOCI, rather than PCA, therefore we do not employ their method in this study.} We inserted three artificial disks (the best-matching model from \citealt{boc32297}) of varying brightness and PA into the raw images. Specifically, the first disk was slightly brighter than the real disk and located \about 90\degrees away; the second was 10$\%$ fainter than the first and located \about 45\degrees away; and the third was 25$\%$ fainter than the first and located \about 135\degrees away. We chose to insert disks of varying SB and PA to better account for the biases inherent in the PCA reduction. 

After recovering both the real and artificial disks (Fig. \ref{fig:fakedisks}), we measured the PCA correction factors\footnote{For two of the artifical disks, the azimuthal separation from the real disk is less than the total parallactic angle rotation (50.84\degrees), which can cause additional unwanted subtraction in the real disk. However, we have determined that this effect is negligible for our data.} as a function of separation from the star by comparing the median counts/s in the same 5 pixel (0\fasec 106) by 5 pixel boxes measured on the PCA-processed artificial disks with the equivalent measurements on the pure, unaltered, noiseless model disk images. We recorded the ratio of these two numbers as a function of distance from the star and PA, then we averaged all the values together at each separation. Fig. \ref{fig:pcabias} shows these correction factors.

With all correction factors computed as a function of separation from the star for each wavelength, we used Eq. \ref{eqn:corrections} to correct the measured SB to the true SB of the disk.


\bibliographystyle{apj}
\bibliography{ms}


\end{document}